\documentclass[journal,a4paper]{IEEEtran}
\usepackage{cite}
\def\e{\begin{equation}}
	\def\f{\end{equation}}

\def\.{\cdot}
\def\x0{\mathbf{x}_0}
\def\y0{\mathbf{y}_0}
\def\z0{\mathbf{z}_0}

\usepackage{gensymb}

\usepackage{amssymb}
\ifCLASSINFOpdf
\usepackage[pdftex]{graphicx}
\else
  \usepackage[dvips]{graphicx}
\fi

\usepackage{caption}

\usepackage{epstopdf}

\usepackage{placeins}

\usepackage[cmex10]{amsmath}
\DeclareMathOperator{\arccot}{arccot}
\usepackage{array}
 \usepackage[caption=false,font=footnotesize]{subfig}

\usepackage{fixltx2e}
\begin{document}
\title{Multifunctional \iffalse Multifrequency \fi Cascaded Metamaterials: \\
Integrated Transmitarrays}

\author{Amr~A.~Elsakka, Viktar~S.~Asadchy, Ihar~A.~Faniayeu, Svetlana~N.~Tcvetkova,
        and~Sergei~A.~Tretyakov,~\IEEEmembership{Fellow,~IEEE}% <-this % stops a space
\thanks{A.~A.~Elsakka, S.~N.~Tcvetkova, and S.~A.~Tretyakov are with Department of Radio Science and Engineering, Aalto University, P.O. 13000, FI-00076, Aalto, Finland (e-mail: amr.elsakka@aalto.fi, svetlana.tcvetkova@aalto.fi).

V.~S.~Asadchy is with the Department of Radio Science and Engineering, Aalto University, FI-00076 Aalto, Finland, and also with Department of General Physics, Francisk Skorina Gomel State University, 246019 Gomel, Belarus (e-mail: viktar.asadchy@aalto.fi).

I.~A.~Faniayeu is with the Research Institute of Electronics, Shizuoka University 3-5-1 Johoku, Naka-ku, Hamamatsu 432-8011, Japan, and also with Department of General Physics, Francisk Skorina Gomel State University, 246019 Gomel, Belarus 
%(e-mail: ).
  %
%; sergei.tretyakov@aalto.fi 
).}% <-this % stops a space

%\thanks{Manuscript received April 19, 2005; revised September 17, 2014.}
}

\maketitle

\begin{abstract}
%Engineering reflection and transmission of electromagnetic waves is very important in a wide range of applications; that is why there is always a continuous need for new and more  efficient solutions. Here, design and realization of a single-layer meta-transmitarray (metasurface) which provides a certain functionality in a narrow frequency band while remains reflectionless (transparent) outside of the operational frequency band is presented. Realizing such transmitarrays would allow integrating several metasurfaces in composite structures that perform different functionalities at different frequencies, opening a door for various new applications. In this paper we provide a theoretical analysis for alternative possible ways to realize such a metasurface. We also introduce two designs for metasurfaces that demonstrate abilities for wavefront shaping and anomalous refraction. We numerically simulate two proposed designs as well as measure one of them experimentally. Moreover, we propose a promising approach to design multifunctional cascaded metasurfaces that provide different operations at different frequencies.

Control of electromagnetic waves using engineered materials is very important in a wide range of applications, therefore there is always a continuous need for new and more efficient solutions. Known natural and artificial materials and surfaces provide a particular  functionality in the frequency range they operate but cast a ``shadow'' and produce reflections at other frequencies. Here, we introduce a concept of multifunctional engineered materials that possess different predetermined functionalities at different frequencies. Such response can be accomplished by cascading metasurfaces (thin composite layers) that are designed to perform a single operation at the desired frequency and are transparent elsewhere. Previously, out-of-band transparent metasurfaces  for control over reflection and absorption were proposed. In this paper, to complete the full set of functionalities for wave control, we synthesize transmitarrays that tailor transmission in a desired way, being ``invisible'' beyond the operational band. 
%The governing idea behind this design is metasurface inclusions with equal frequency dispersions ..
The designed transmitarrays for wavefront shaping and anomalous refraction are tested numerically and experimentally. 
To demonstrate our concept of multifunctional engineered materials, we have designed a cascade of three metasurfaces that performs three different functions for waves at different frequencies. Remarkably, applied to volumetric metamaterials, our concept can enable a single composite possessing desired multifunctional  response.

\end{abstract}


% \fi
\IEEEpeerreviewmaketitle

\section{Introduction}

\IEEEPARstart{M}{anipulations} of electromagnetic waves in transmission through various structures  has been of fundamental importance in a great number of applications. Through interaction of waves with matter it is possible to control the wave intensity, polarization and propagation direction. The simplest devices for wave control, such as optical lenses and mirrors, have evolved into numerous appliances operating with radiation from radiowaves to ultraviolet:
Fresnel and dielectric lenses \cite{cornbleet}, antenna arrays \cite{reflectarrays1,reflectarrays2}, etc. 

\begin{figure}[b!]
{\par\centering
 \subfloat[]{{\includegraphics[width=0.43\columnwidth]{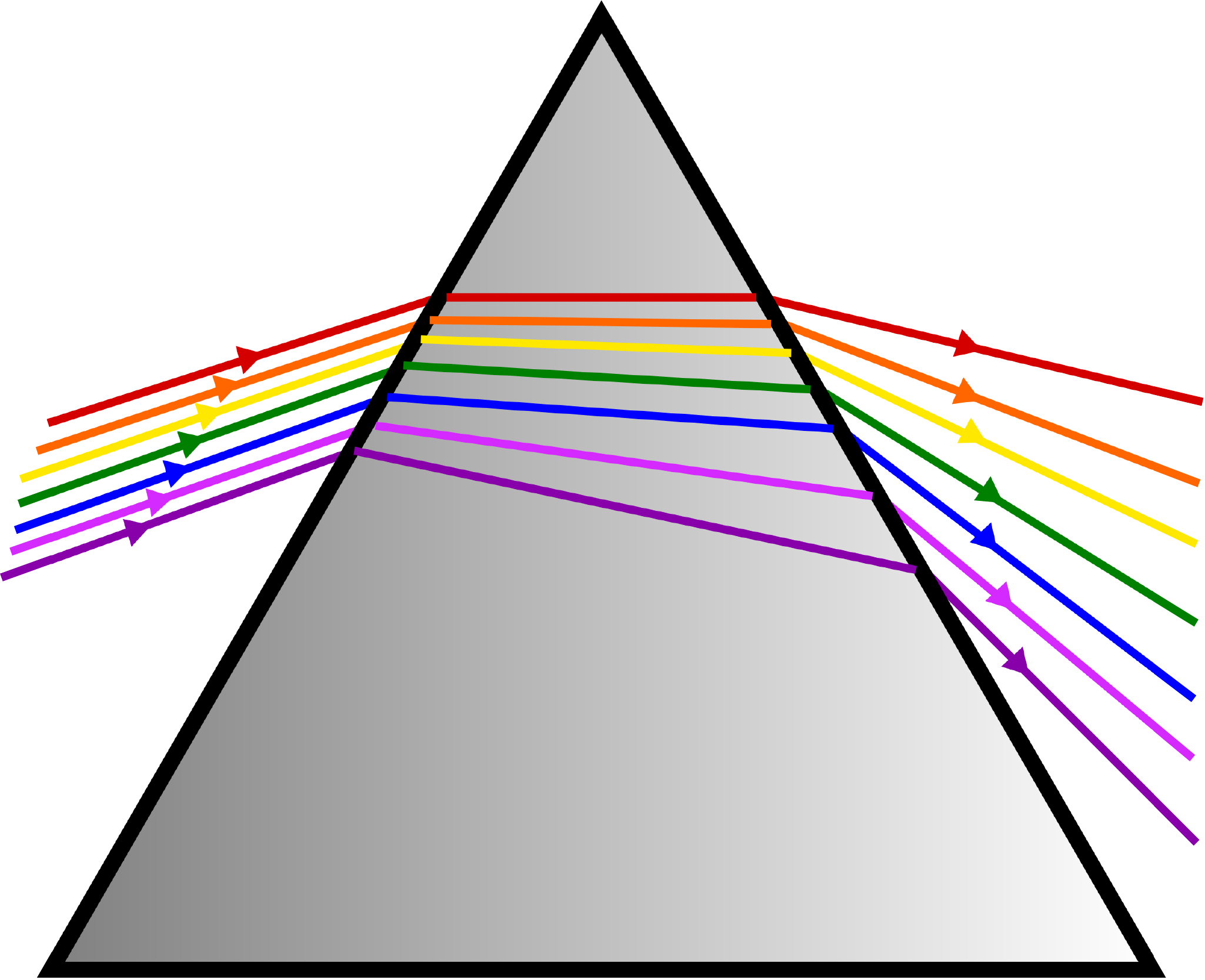} }
  \label{fig:fig1a}} 
 \qquad
 \subfloat[]{{\includegraphics[width=0.40\columnwidth]{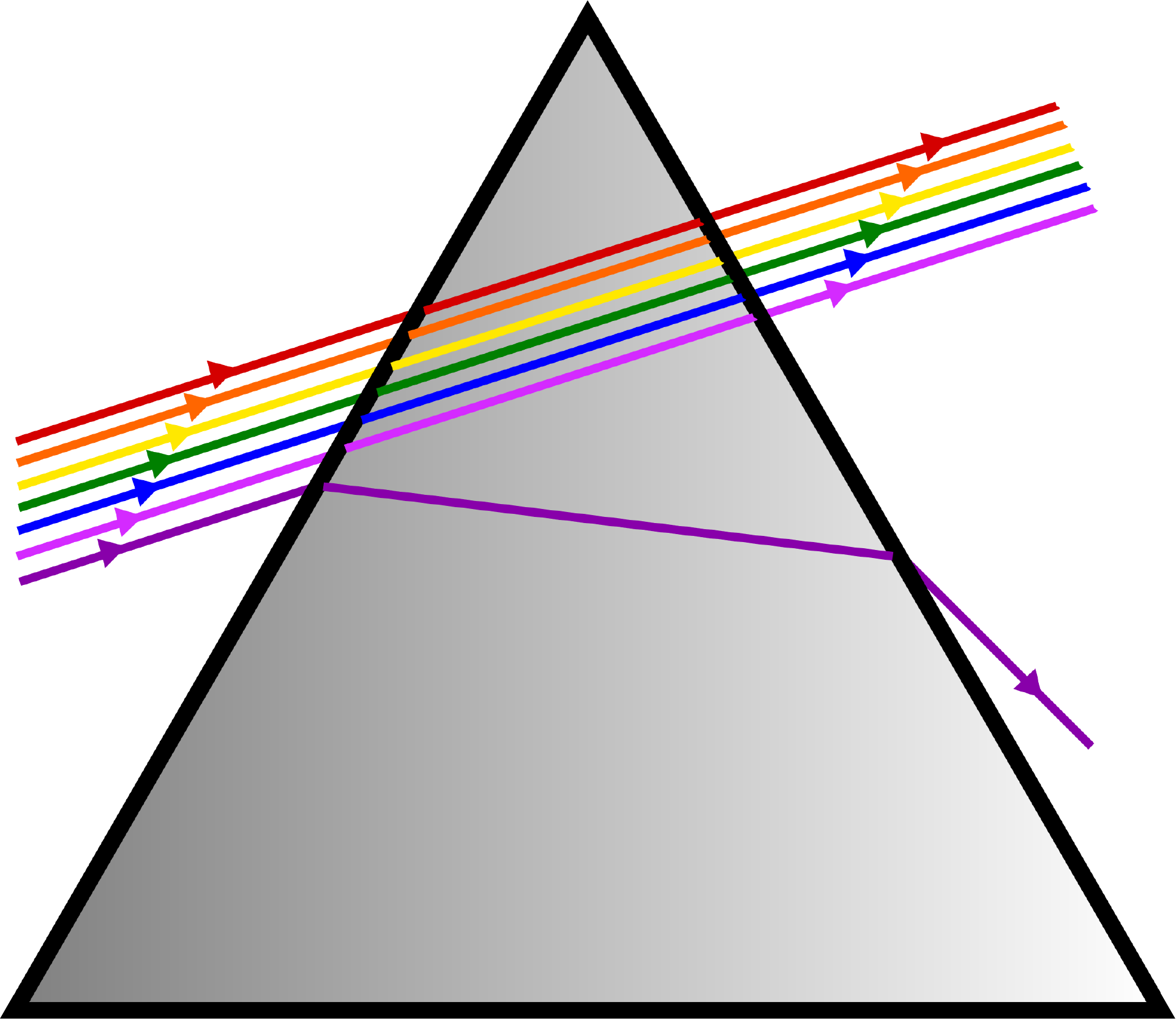} }
  \label{fig:fig1b}} 
\\
\subfloat[]{{\includegraphics[height=2.5cm,width=0.95\columnwidth]{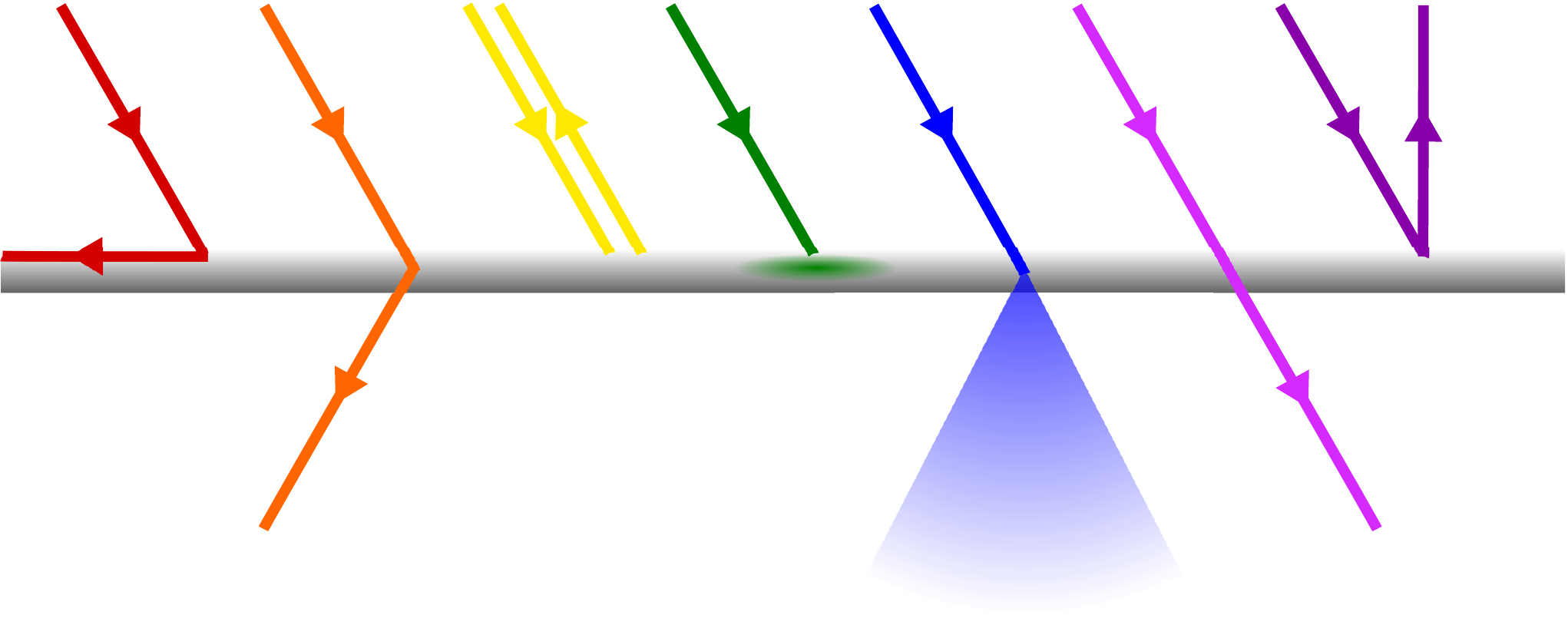} }
  \label{fig:fig1c}} 
 \caption{(a) Light propagation through Newton's prism. The prism refracts the light of all  frequencies. (b) Light propagation through a prism that refracts only light of violet color and does not interact with light of other frequencies. (c) Conceptual illustration of a multifunctional multifrequency composite. Incident light of different colors receives different predetermined response from the composite.
 }
 \label{fig:fig1}}
\end{figure}

Almost all known structures for wave manipulations perform a particular functionality in the frequency range they operate, while being not transparent and casting a ``shadow'' (or creating some disturbance)  at other frequencies.  Figure~\ref{fig:fig1a} illustrates such functionality by the example of Newton's prism. The prism designed for light of one color inevitably  disturbs the paths of light of other colors.
On the other hand, designing structures that  manipulate waves only of specific frequencies (see Fig.~\ref{fig:fig1b}), not interacting with radiation of other frequencies, would enable new exciting opportunities. In particular, such devices performing different functionalities at different frequencies could be cascaded and even combined in one \emph{single} structure (if its constitutional elements are of several different types) [see Fig.~\ref{fig:fig1c}].

%\begin{figure}[h]
%\centering
% \subfigure[]{
%   \epsfig{file=fig2a.eps, width=0.45\columnwidth}
%   \label{ris:fig1a} }
%  \subfigure[]{
%   \epsfig{file=fig2b.eps, width=0.45\columnwidth}
%   \label{ris:fig1b} }
%\caption{}
%\label{ris:fig1}
%\end{figure}

Even arrays of very small and non-resonant elements inevitably reflect and absorb electromagnetic waves, and here we focus on new design solutions which allow minimization of such parasitic interactions everywhere except the desired operational band. 
To the best of our knowledge, such frequency-selective control of electromagnetic radiation has been explored only in artificial composites manipulating reflection \cite{FuncMetamirror,FuncMetamirror2} and absorption \cite{PRX} of incident waves. To complete the full set of functionalities for wave control, it is necessary to design a frequency-selective transmitter, i.e. a structure that tailors transmitted fields of incident waves of desired frequencies and passes through the others. Such ``transmitters'' can be integrated into a cascade of different devices independently performing  multifunctional multifrequency operations.
Obviously, conventional optical and microwave lenses cannot take on the role of such a structure. Another candidate could be transmitarray antennas (also called array
lenses) invented several decades ago \cite{transmitarray1,transmitarray2}. 
They significantly extended our opportunities for wave control, enabling wavefront shaping and beam scanning. Conventional transmitarray antennas incorporate a ground plane with the receiving and transmitting antenna arrays on its sides connected by matched cables. Therefore, transmitarray antennas cannot pass through the incident radiation at the frequencies beyond their bands, casting a shadow.

Recently, there have been considerable interest and progress in manipulation of  electromagnetic waves using metamaterials  \cite{capasso,shalaev,TailoringWavefrontsPRL,elefth,alu,caloz,PassiveReflectionless,polarization,passive,elefthreview,elefth2,ours}. In this scenario, the structure represents a  composite comprising a two-dimensional array of sub-wavelength elements (so-called \emph{metasurface}). The elements are electrically and magnetically polarizable so that the dipole moments induced in each element form Huygens' pairs. Therefore, each element does not scatter in the backward direction (producing zero reflection), while it radiates waves with the prescribed phase and amplitude in the forward direction. The forward scattered waves from the elements together with the incident wave form the transmitted wave. 
It should be noted that in all known transmitarray metasurfaces the structural elements constitute reflectionless Huygens' sources only inside a narrow frequency band. Beyond this band reflections appear because of prevailing excitation of either electric or magnetic dipole. This is due to the fact that these elements have different frequency dispersions of the electric and magnetic dipole modes (see \cite{PRX}). 

In order to design a structure that transforms waves but is invisible for the incident radiation outside of the operational band (see Fig.~\ref{fig:fig1b}), its elements should be designed in such a way that the electric and magnetic dipole moments, induced in them, are balanced (have equal amplitudes) at all practically relevant frequencies. This implies that both dipole responses should be created by excitation of the same resonant mode (the dipole  moments are formed by the same current distribution in the element). Such regime is possible only if each element consists of a \emph{single} conductive wire or strip \cite{PRX}. These wire elements inevitably possess bianisotropic properties \cite{bian}. 
Interestingly, being narrowband, the single-wire Huygens' elements do not produce reflections in a very broad range of frequencies. In the earlier work, such scenario was realized only for absorbers \cite{PRX}, but not for transmitarrays.

In this paper, we synthesize a uniaxial (isotropic in its plane) low-loss reciprocal metasurface that transforms the wavefront of incident waves in a desired manner (in transmission) at a desired frequency, remaining transparent in a wide frequency band. 
We analyse all possible scenarios of realization of such a metasurface and determine the unique requirement for the electromagnetic response of its elements. We design and test two synthesized metasurfaces that demonstrate their abilities for wavefront shaping and anomalous refraction. Moreover, we propose a promising approach to design multifunctional cascaded metasurfaces that provide different operations at different frequencies (similarly to the conceptual example in Fig.~\ref{fig:fig1c}).  Our approach, generally, can be extended to volumetric metamaterials.
We find design solutions for integrated metasurfaces that provide three basic functions such as full control over reflection, absorption and transmission properties.  Based on these metasurfaces, one can synthesize arbitrary cascaded composites for general multifunctional  wave manipulation.

\section{Controlling the phase and amplitude of transmission with single-wire Huygens' elements}
%Previously, it was shown that limited manipulation of waves transmitted through a thin metasurface can be accomplished due to specifically designed phase gradient over the metasurface plane \cite{capasso,shalaev,grady}. The phase gradient is achieved by precise adjustment of the phases of transmitted waves from each metasurface inclusion. Although, this approach have enabled simple realization of transmitarrays even at optical frequencies, it suffers from very low efficiency (less than 25\%) and cannot provide control of polarization of the transmitted waves (in fact, it suffers from uncontrollable polarization rotation by $90^\circ$). Subsequently, another approach based on generalized boundary conditions was proposed by several researchers \cite{Tailoring Wavefronts PRL,elefth,alu,caloz}. It provided high efficient operation (nearly 80\%) and manipulation of polarization \cite{polarization}. However, even this approach could not ensure ideal performance \cite{passive,elefthreview} due to the fact that it forbids the bianisotropic response in the metasurface. A complete approach that offers the ideal and full control over transmission was newly introduced \cite{elefth2,ours} and considered the  scenario where the metasurface can have general bianisotropic properties.

Manipulation of waves transmitted through a thin metasurface can be accomplished due to specifically designed phase gradient over the metasurface plane (e.g., \cite{capasso,shalaev,grady}). The phase gradient can be achieved by precise adjustment of the phases of transmitted waves from each metasurface inclusion. To adjust the phases for each inclusion, we utilize so-called locally uniform homogenization approach, i.e. we tune an individual inclusion, assuming that it is located in an array with a uniform phase distribution. An array of such individually adjusted inclusions possesses nearly the required non-uniform phase distribution.
Therefore, it is important to design individual inclusions so that uniform arrays formed by them transmit incident waves, conserving its  amplitude but changing its phase by a specific value $\phi$ (different for each inclusion) that belongs to the interval from 0 to $2\pi$. Next, we examine all possible scenarios of designing metasurface elements that satisfy these conditions.

Let us  consider a reciprocal metasurface (the same transmission properties from both sides) as a two-dimensional periodic array of sub-wavelength bianisotropic inclusions polarizable electrically and magnetically. The ability of the inclusions to get polarized in the external electric and magnetic fields is described, respectively, by the effective polarizability dyadics $\overline{\overline{\widehat{\alpha}}}_{\rm ee}=(\overline{\overline{\widehat{\alpha}}}_{\rm ee})^T$ and $\overline{\overline{\widehat{\alpha}}}_{\rm mm}=(\overline{\overline{\widehat{\alpha}}}_{\rm mm})^T$, where $T$ denotes the transpose operation.
Bianisotropy implies that the electric (magnetic) field of the incident wave can also produce magnetic (electric) polarization in the inclusions. This effect is often called magnetoelectric coupling and can be characterized by the magnetoelectric polarizability dyadic $\overline{\overline{\widehat{\alpha}}}_{\rm me}$, which for reciprocal structures is connected with the electromagnetic polarizability as $\overline{\overline{\widehat{\alpha}}}_{\rm me}=-(\overline{\overline{\widehat{\alpha}}}_{\rm em})^T$ \cite{bian}. 
Considering the uniaxial symmetry of the metasurface, it is convenient to represent the polarizability dyadics in the following form:
\begin{equation}
\begin{array}{c}
\overline{\overline{\widehat{\alpha}}}_{\rm ee}=\widehat{\alpha}_{\rm ee}^{\rm co} \overline{\overline{I}}_{\rm t}+\widehat{\alpha}_{\rm ee}^{\rm cr} \overline{\overline{J}}_{\rm t},\qquad \displaystyle
\overline{\overline{\widehat{\alpha}}}_{\rm mm}=\widehat{\alpha}_{\rm mm}^{\rm co} \overline{\overline{I}}_{\rm t}+\widehat{\alpha}_{\rm mm}^{\rm cr} \overline{\overline{J}}_{\rm t},  \\\vspace*{.1cm}\displaystyle
\overline{\overline{\widehat{\alpha}}}_{\rm em}=\widehat{\alpha}_{\rm em}^{\rm co} \overline{\overline{I}}_{\rm t}+\widehat{\alpha}_{\rm em}^{\rm cr} \overline{\overline{J}}_{\rm t},\qquad\displaystyle
\overline{\overline{\widehat{\alpha}}}_{\rm me}=\widehat{\alpha}_{\rm me}^{\rm co} \overline{\overline{I}}_{\rm t}+\widehat{\alpha}_{\rm me}^{\rm cr} \overline{\overline{J}}_{\rm t},\displaystyle
\end{array}\label{eq:uniaxial1}
\end{equation}
where $\overline{\overline{I}}_{\rm t}$ and $\overline{\overline{J}}_{\rm t}$ are the transverse unit and vector-product dyadics, respectively, and the indices $\rm co$ and $\rm cr$ refer to the symmetric and antisymmetric parts of the corresponding dyadics. 
Taking into account the reciprocity of the metasurface, equations (\ref{eq:uniaxial1}) can be rewritten as
\begin{equation}
\begin{array}{c}
\overline{\overline{\widehat{\alpha}}}_{\rm ee}=\widehat{\alpha}_{\rm ee}^{\rm co} \overline{\overline{I}}_{\rm t},\qquad \displaystyle
\overline{\overline{\widehat{\alpha}}}_{\rm mm}=\widehat{\alpha}_{\rm mm}^{\rm co} \overline{\overline{I}}_{\rm t},  \\\vspace*{.1cm}\displaystyle
\overline{\overline{\widehat{\alpha}}}_{\rm em}=\widehat{\alpha}_{\rm em}^{\rm co} \overline{\overline{I}}_{\rm t}+\widehat{\alpha}_{\rm em}^{\rm cr} \overline{\overline{J}}_{\rm t},\qquad\displaystyle
\overline{\overline{\widehat{\alpha}}}_{\rm me}=-\widehat{\alpha}_{\rm em}^{\rm co} \overline{\overline{I}}_{\rm t}+\widehat{\alpha}_{\rm em}^{\rm cr} \overline{\overline{J}}_{\rm t}.\displaystyle
\end{array}\label{eq:uniaxial2}
\end{equation}

Assuming that the incident wave impinges on the uniaxial metasurface normally along the $-z$-axis, the electric fields of the reflected and transmitted plane waves from the metasurface are given by \cite{niemi}
\begin{equation}
			\displaystyle
\mathbf{E}_{\rm r}=  -\frac{j\omega}{2S}\left[\eta_{0}\widehat{\alpha}^{\rm co}_{\rm ee}+2 \widehat{\alpha}^{\rm cr}_{\rm em}-\frac{1}{\eta_{0}}\widehat{\alpha}^{\rm co}_{\rm mm}\right]\cdot\mathbf{E}_{\rm inc}, \vspace*{.2cm}\\ \displaystyle
		\label{eq:q21}
\end{equation}
\begin{equation}
			\displaystyle
\mathbf{E}_{\rm t}=  \left[\left( 1-\frac{j\omega}{2S}\big[\eta_{0}\widehat{\alpha}^{\rm co}_{\rm ee}+\frac{1}{\eta_{0}}\widehat{\alpha}^{\rm co}_{\rm mm}\big] \right)\overline{\overline{I}}_{\rm t}   +\frac{j\omega}{S} \widehat{\alpha}^{\rm co}_{\rm em}\overline{\overline{J}}_{\rm t}\right]\cdot\mathbf{E}_{\rm inc}, 
\label{eq:q22}
\end{equation}
where $\omega$ is the angular frequency, $S$ is the area of the array unit cell, and $\eta_{0}$ is the free-space wave impedance.

\begin{figure}[b!]
\centering
 \subfloat[]{{\includegraphics[width=0.45\columnwidth]{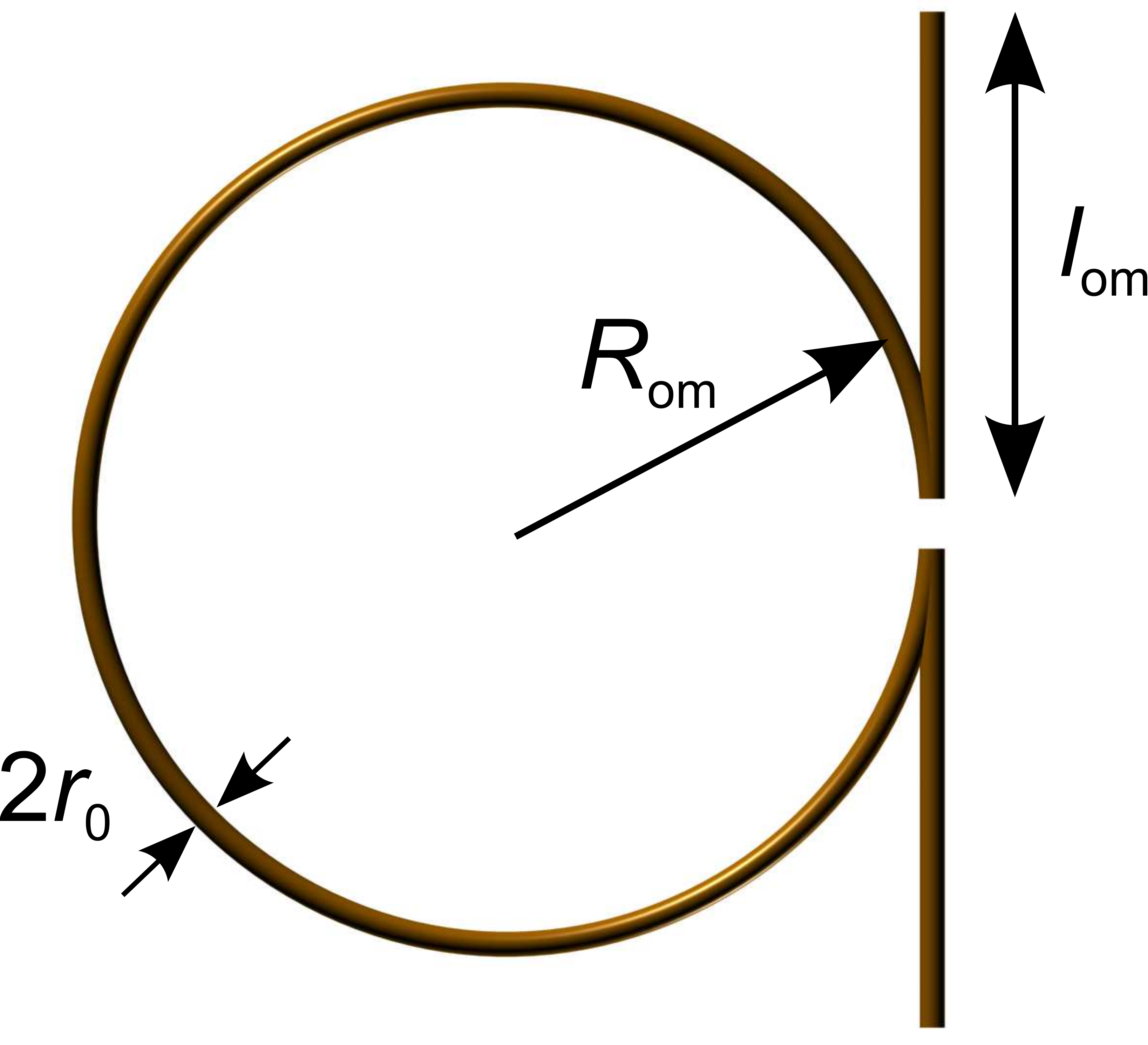} }
  \label{fig:fig2a}} 
 \subfloat[]{{\includegraphics[width=0.45\columnwidth]{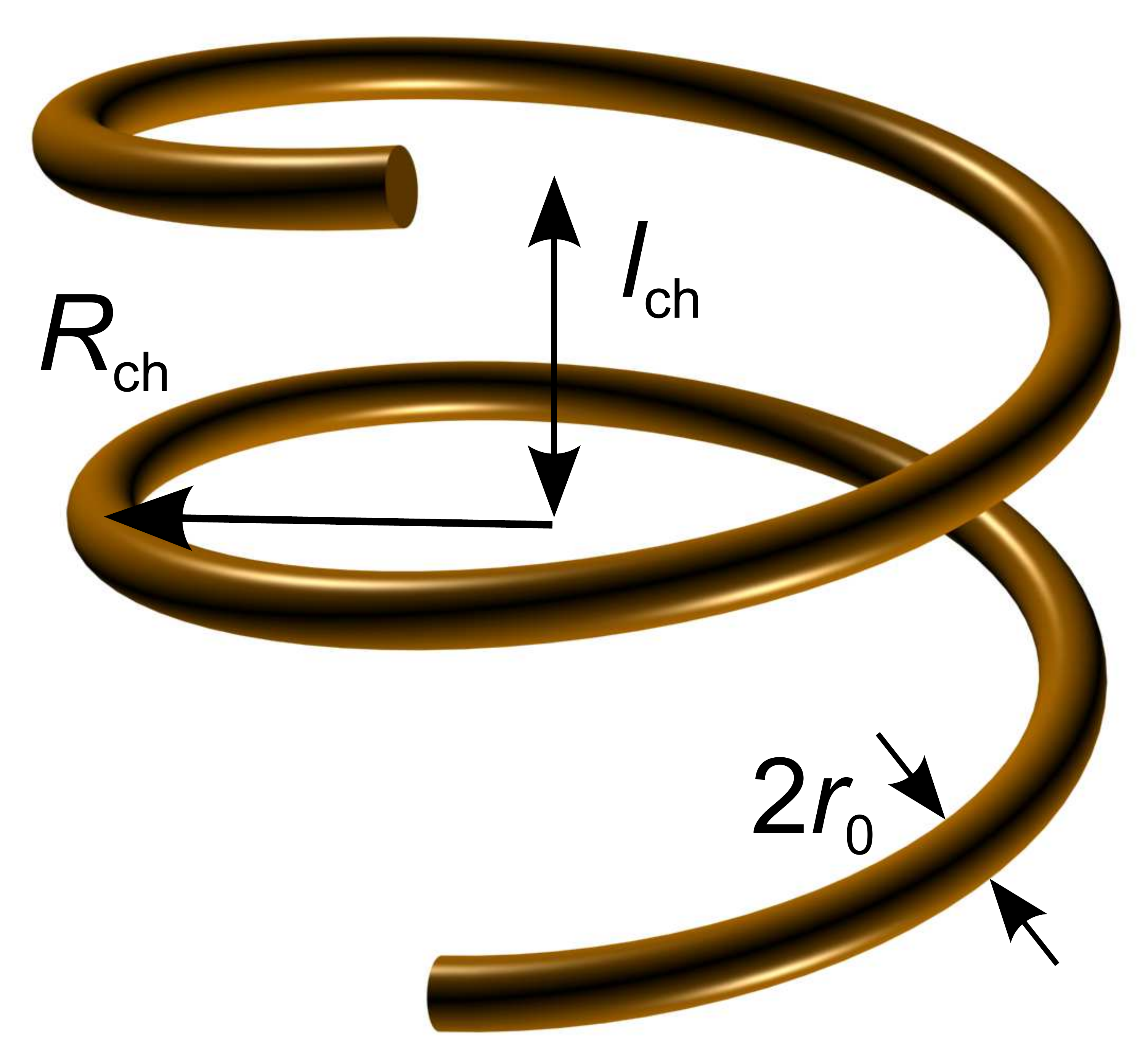} }
  \label{fig:fig2b}} 
 \caption{Examples of bianisotropic single-wire inclusions. (a) Omega inclusion. (b) Chiral inclusion. }
 \label{fig:fig2}
\end{figure}

As discussed in the introduction, to realize broadband reflectionless regime, the metasurface elements must be bianisotropic single-wire inclusions (see examples in Fig.~\ref{fig:fig2}). 
In the literature, bianisotropy is usually classified to two classes: chiral class with symmetric electromagnetic dyadic ($\widehat{\alpha}_{\rm em}^{\rm cr}=0$) and omega class, when the dyadic is   antisymmetric   ($\widehat{\alpha}_{\rm em}^{\rm co}=0$) \cite{bian,Totalabsorption}.
Based on this classification, for the sake of clarity, we consider this two cases separately. 

\subsection{Arrays with  single-wire bianisotropic omega elements}
For a uniform array of single-wire omega inclusions  (see Fig.~\ref{fig:fig2a}) the following relation between  the effective polarizabilities of each inclusion holds \cite{bian,Totalabsorption}:
\begin{equation}
\widehat{\alpha}^{\rm co}_{\rm ee}\widehat{\alpha}^{\rm co}_{\rm mm}=-\widehat{\alpha}^{\rm cr}_{\rm em}\widehat{\alpha}^{\rm cr}_{\rm me}=-(\widehat{\alpha}^{\rm cr}_{\rm em})^2.\label{eq:q3}
\end{equation}
%\begin{equation}
%\Rightarrow\hat{\alpha}^{\rm cr}_{\rm em}=\sqrt{-\hat{\alpha}^{\rm co}_{\rm ee}\hat{\alpha}^{\rm co}_{\rm mm}}
%\end{equation}
Substituting ${\widehat\alpha^{\rm cr}_{\rm em}}$ from (\ref{eq:q3}) in (\ref{eq:q21}), we find the fields of reflected waves from the omega metasurface:
\begin{equation}
\displaystyle
\mathbf{E}_{\rm r}=  -\frac{j\omega}{2S}\left[\eta_{0}\widehat{\alpha}^{\rm co}_{\rm ee}+2\sqrt{-\widehat{\alpha}^{\rm co}_{\rm ee}\widehat{\alpha}^{\rm co}_{\rm mm}} -\frac{1}{\eta_{0}}\widehat{\alpha}^{\rm co}_{\rm mm}\right]\cdot\mathbf{E}_{\rm inc}.\vspace*{.2cm}\\ \displaystyle
 \label{eq:q4}
\end{equation}
Thus, the condition of zero reflection essential for our transmitarray ($\mathbf{E}_{\rm r}=0$)  implies a limitation on the effective polarizabilitities:
%\begin{equation}
%\rm\eta_{o}\hat{\alpha}^{co}_{ee}+2\sqrt{-\hat{\alpha}^{co}_{ee}\hat{\alpha}^{co}_{mm}}-\frac{1}{\eta_{o}}\hat{\alpha}^{co}_{mm}=0
%\label{eq:q5}
%\end{equation}
%Solving this quadratic equation will lead to:
\begin{equation}
\widehat{\alpha}^{\rm co}_{\rm ee}=-\frac{1}{\eta^{2}_{0}}\widehat{\alpha}^{\rm co}_{\rm mm}.
\label{eq:q5}
\end{equation}
This limitation on the effective polarizabilities (which takes into account interactions between the inclusions) leads to a corresponding limitation for the individual polarizabilities (modelling the properties of an individual particle in free space) \cite{Totalabsorption}: ${\alpha}^{\rm co}_{\rm ee}=(-1/\eta^{2}_{0})\,{\alpha}^{\rm co}_{\rm mm}$. This condition, obviously, cannot be satisfied with passive inclusions. Indeed, the opposite signs of the electric and magnetic polarizabilities imply that their imaginary parts have the opposite signs. This scenario corresponds to the case of a passive-active pair of dipole moments. Furthermore, one can see from (\ref{eq:q22}) [assuming $\widehat{\alpha}_{\rm em}^{\rm co}=0$] and (\ref{eq:q5}) that in this case the phase of the transmitted wave through the metasurface is always equal to that of the incident wave ($\mathbf{E}_{\rm t}=\mathbf{E}_{\rm inc}$). Thus, it is impossible to synthesize a transmitarray with the desired properties using single-wire omega elements. 
%Interestingly, omega elements can be used to realize

\subsection{Arrays with  single-wire bianisotropic chiral elements}
Likewise, effective polarizabilities of chiral single-wire inclusions (see Fig.~\ref{fig:fig2b}) in a uniform array are related to one another as follows \cite{bian,Totalabsorption}:
\begin{equation}
 \widehat{\alpha}^{\rm co}_{\rm ee}\widehat{\alpha}^{\rm co}_{\rm mm}=\widehat{\alpha}^{\rm co}_{\rm em}\widehat{\alpha}^{\rm co}_{\rm me}=-(\widehat{\alpha}^{\rm co}_{\rm em})^2
\label{eq:q6}
\end{equation}
%\begin{equation}
%\Rightarrow\widehat{\alpha}^{\rm co}_{\rm em}=\mp\sqrt{\widehat{\alpha}^{\rm co}_{\rm ee}\widehat{\alpha}^{\rm co}_{\rm mm}}
%\label{eq:q7}
%\end{equation}
One can see from (\ref{eq:q21}) that in the case of a chiral metasurface ($\widehat{\alpha}_{\rm em}^{\rm cr}=0$), the condition of zero reflection ($\mathbf{E}_{\rm r}=0$) simply requires the balanced electric and magnetic dipoles $\widehat{\alpha}^{\rm co}_{\rm ee}=(1/\eta^{2}_{0})\,\widehat{\alpha}^{\rm co}_{\rm mm}$ of each metasurface inclusion. Taking this result into account and combining with relation (\ref{eq:q6}), the transmitted fields  through the chiral metasurface (\ref{eq:q22}) can be written as
\begin{equation}
	\mathbf{E}_{\rm t}=\left[\left(1-\frac{j\omega}{S}\eta_{0}\widehat{\alpha}^{\rm co}_{\rm ee}\right)\overline{\overline{I}}_{\rm t} \pm  \frac{\omega}{S}\eta_{0}\widehat{\alpha}^{\rm co}_{\rm ee}\overline{\overline{J}}_{\rm t}\right]\cdot\mathbf {E}_{\rm inc},
	\label{eq:q10} 
\end{equation}
where the upper and lower signs correspond to chiral inclusions with the right and left handedness, respectively.

From (\ref{eq:q10}) it is seen that, generally, the polarization of the wave transmitted through a chiral transmitarray is different from that of the incident wave. 
In designs of conventional transmitarrays almost always it is assumed that the polarization of the wave passing through a transmitarray does not change. However, in many applications polarization-plane rotation of transmitted waves (in focusing arrays designed for circularly polarized waves, for example), is acceptable. Thus, it is important to consider also the case when the transmitarray transforms the  incident wave polarization, since if there is no requirement for keeping the polarization constant, there is more design freedom in transmitarrays realizations.
Therefore, we  look for a solution for the transmitted field in the most general form of elliptical polarization:
\begin{equation}
	\mathbf{E}_{\rm t}=(T_{\rm co}\overline{\overline{I}}_{\rm t}+ T_{\rm cr}e^{j\Delta\phi}\overline{\overline{J}}_{\rm t})e^{j\phi}\cdot\mathbf {E}_{\rm inc},
	\label{eq:q11}
\end{equation}
where $\Delta\phi$ is the phase difference between the two orthogonal components of the elliptically polarized transmitted field, $T_{\rm co}$ and $T_{\rm cr}$ are the semi-major and semi-minor axes of the polarization ellipse (real values), and $\phi$ is the phase shift between the  incident wave (assumed to be linearly polarized) and the elliptically polarized transmitted wave.

Comparing (\ref{eq:q10}) and~(\ref{eq:q11}), we find
\begin{equation}
	T_{\rm co}=|1-\frac{j\omega}{S}\eta_{0}\widehat{\alpha}^{\rm co}_{\rm ee}|,
	\qquad 
	\mathbf \phi=\angle(1-\frac{j\omega}{S}\eta_{0}\widehat{\alpha}^{\rm co}_{\rm ee}),
	\label{eq:q12} 
\end{equation}
\begin{equation}
	T_{\rm cr}=|\frac{\omega}{S}\eta_{0}\widehat{\alpha}^{\rm co}_{\rm ee}|,
	\qquad
	\mathbf \phi+\Delta\phi=\angle(\pm \frac{\omega}{S}\eta_{0}\widehat{\alpha}^{\rm co}_{\rm ee}), 
	\label{eq:q13}  
\end{equation}
where symbol $\angle$ denotes the phase angle of the exponential representation of a complex number. 

From the energy conservation in lossless metasurfaces it follows that $T^{2}_{\rm co}+T^{2}_{\rm cr}=1$, which connects the real and imaginary parts of the electric polarizability of each unit cell:
\begin{equation}
	\Re\{\widehat{\alpha}^{\rm co}_{\rm ee}\}=\pm\sqrt{-\Im\{\widehat{\alpha}^{\rm co}_{\rm ee}\}\left(\frac{S}{\omega \eta_{0}}+\Im\{\widehat{\alpha}^{\rm co}_{\rm ee}\}\right)}.
	\label{eq:q14}
\end{equation}
Using (\ref{eq:q14}), we can rewrite (\ref{eq:q12}) and (\ref{eq:q13}) as
\begin{equation}
	T_{\rm co}=\sqrt{1+\frac{\omega}{S}\eta_{0} \Im\{\widehat{\alpha}^{\rm co}_{\rm ee}\}},
	\qquad 
	T_{\rm cr}=\sqrt{-\frac{\omega}{S}\eta_{0} \Im\{\widehat{\alpha}^{\rm co}_{\rm ee}\}},
	\qquad
	\label{eq:q15} 
\end{equation}
\begin{equation}
	\phi=\mp\arccot{\sqrt{-\frac{S}{\omega \eta_{0} \Im\{\widehat{\alpha}^{\rm co}_{\rm ee}\}}-1}},
	\label{eq:q16}  
\end{equation}
and $\Delta \phi=0$.

It should be noted that in order to achieve the maximum efficiency, all the elements of the transmitarray must radiate waves of the same polarization, ensuring constructive interference.  This implies that the polarization parameters $T_{\rm co}$ and $T_{\rm cr}$ should be equal for all the elements. Therefore, from (\ref{eq:q15}) one can see that the imaginary part of the polarizability $\Im\{\widehat{\alpha}^{\rm co}_{\rm ee}\}$ must be the same for all the elements. Evidently, in this case, from (\ref{eq:q16}) we see that the phases of the transmitted waves from each element $\phi$ are equal and cannot be adjusted arbitrarily. This fact forbids designing efficient transmitarrays for wavefront control with single-wire chiral inclusions. 
%Thus, we can conclude that wavefront shaping with polarization transformation cannot be achieved using elements consisting of a single wire. 

%The second condition shows that $T_{co}$ and $T_{cr}$ are of fixed value, from~(\ref{eq:qaa}) it's clear that:
%\begin{equation}
%	\Re\{\hat{\alpha}^{co}_{ee}\}^{2}+\Im\{\hat{\alpha}^{co}_{ee}\}^{2}=constant
%		\label{eq:qd}
%\end{equation}
%Considering both~(\ref{eq:qc}) and~(\ref{eq:qd}) it's clear the the real and imaginary parts of $\hat{\alpha}^{co}_{ee}$ are constants.
%And ~(\ref{eq:qb}) shows the $\phi$ is also constant which means that we can't control the transmitted wave's phase, and that is coming from the second restriction. 
\subsection{Non-bianisotropic arrays with single-wire elements}
In the previous sections it was shown that design of a transmitarray which is ``invisible'' beyond its operational band requires the use of bianisotropic single-wire inclusions. On the other hand, it was demonstrated that bianisotropic arrays of single-wire inclusions do not provide full phase control from 0 to $2\pi$. The only solution to overcome these two contradictory statements is designing a transmitarray whose each unit cell consists of bianisotropic inclusions, being in overall  not bianisotropic. This situation is possible if the bianisotropic effects of the inclusions in a single unit cell are mutually compensated. To realize it, one can compose a unit cell of inclusions with the opposite (by sign) bianisotropy parameters. Therefore, there can be two  different but equivalent scenarios: a unit cell consists of chiral inclusions with left and right handedness \cite{semchenko,PRX} and a unit cell consists of oppositely oriented omega inclusions \cite{Balmakou}. In both these cases the bianisotropic effects are completely compensated and the unit cell behaves as a pair of orthogonal electric and magnetic dipoles. However, in contrast to the well known unit cells consisting of a split ring resonator and a continuous wire \cite{SRR1,SRR2}, this anisotropic unit cell made of bianisotropic elements is reflectionless and ``invisible'' over a very broad frequency range.

The field expressions (\ref{eq:q21}) and (\ref{eq:q22}) for the array of single-wire inclusions with compensated bianisotropy we rewrite as
\begin{equation}
\displaystyle
\mathbf{E}_{\rm r}=  -\frac{j\omega}{2S}\left[\eta_{0}\widehat{\alpha}^{\rm co}_{\rm ee}-\frac{1}{\eta_{0}}\widehat{\alpha}^{\rm co}_{\rm mm}\right]\cdot\mathbf{E}_{\rm inc}, \vspace*{.2cm}\\ \displaystyle
		\label{eq:q31}
\end{equation}
\begin{equation}
\displaystyle
\mathbf{E}_{\rm t}=  \left[ 1-\frac{j\omega}{2S}\big(\eta_{0}\widehat{\alpha}^{\rm co}_{\rm ee}+\frac{1}{\eta_{0}}\widehat{\alpha}^{\rm co}_{\rm mm}\big) \right]\cdot\mathbf{E}_{\rm inc},
\label{eq:q32}
\end{equation}
where reflection from the metasurface is suppressed, only if the dipole moments of the unit cells are balanced  $\widehat{\alpha}^{\rm co}_{\rm ee}=(1/\eta^{2}_{0})\,\widehat{\alpha}^{\rm co}_{\rm mm}$.

\begin{figure}[h]
\centering
 \subfloat[]{{\includegraphics[width=0.49\columnwidth]{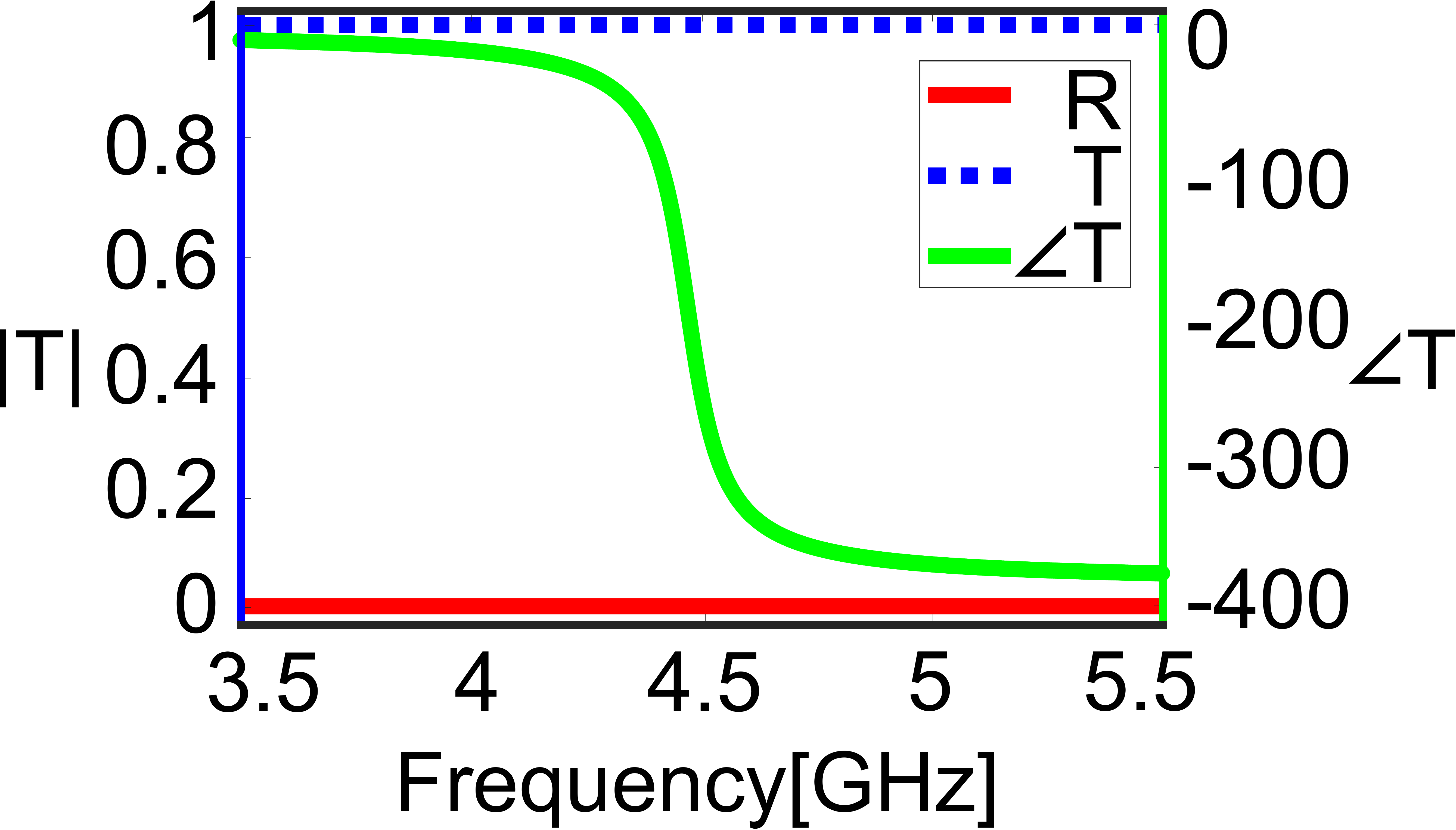} }
  \label{fig:fig3a}} 
   \subfloat[]{{\includegraphics[width=0.49\columnwidth]{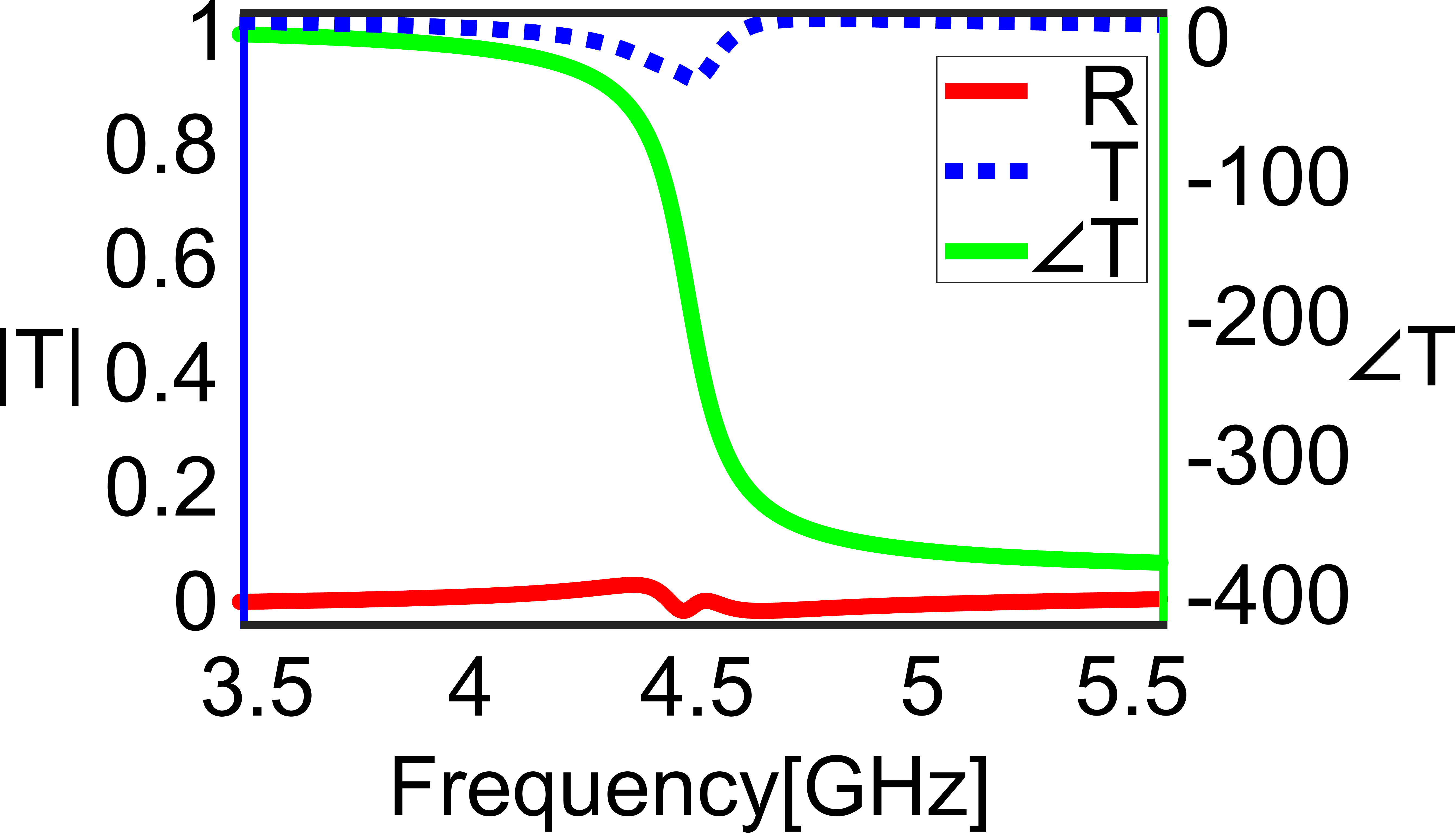} }
  \label{fig:fig3b}} 
 \caption{Reflectance $R$, transmittance $T$ and the phase of transmission $\angle T$ from a periodic array of single-wire inclusions. (a) Theoretical model of a lossless anisotropic metasurface. (b) Numerical results for the  structure depicted in Fig.~\ref{fig:fig4}.   }
 \label{fig:fig3}
\end{figure}

Assuming that the effective polarizabilities of lossless \mbox{balanced} inclusions in a periodic array can be written as \cite{analyticalmodelling}
\begin{equation}
\displaystyle
\frac{1}{\eta_{0}\widehat{\alpha}^{\rm co}_{\rm ee}} = \frac{1}{\widehat{\alpha}^{\rm co}_{\rm mm}/\eta_{0}}=\Re \left\{ \frac{1}{\eta_{0}\alpha^{\rm co}_{\rm ee}} \right\}
+j \frac{\omega}{2S},
\label{eq:phase1}
\end{equation}
one can find from (\ref{eq:q32}) the fields transmitted through the metasurface:
\begin{equation}
\displaystyle
\mathbf{E}_{\rm t}=  \left[ 1-\frac{j\omega}{S}\eta_{0}\widehat{\alpha}^{\rm co}_{\rm ee} \right]\cdot\mathbf{E}_{\rm inc}=e^{\displaystyle -j \phi_{\rm t}} \cdot\mathbf{E}_{\rm inc},
\label{eq:phase2}
\end{equation}
where
\begin{equation}
\displaystyle
\phi_{\rm t}=2 \arctan \left(\frac{\omega \eta_0}{2S}\frac{1}{\Re \left\{
1/\alpha^{\rm co}_{\rm ee}  \right\}
 }\right) .
\label{eq:phase3}
\end{equation}
%\begin{equation}
%\displaystyle
%\phi_{\rm t}=2 \arctan \left( \frac{\omega \eta_0}{2S}\frac{|\widehat{\alpha}^{\rm co}_{\rm ee}|}{\Re \left\{
%\widehat{\alpha}^{\rm co}_{\rm ee} \right\}
% }  \right).
%\label{eq:phase3}
%\end{equation}
Figure~\ref{fig:fig3a} shows the amplitude and phase of the transmitted wave, dictated by (\ref{eq:phase2}) and (\ref{eq:phase3}),  through a uniform anisotropic array of  single-wire inclusions. Here we have assumed that the real part of the individual polarizability of the unit cell has Lorentzian dispersion $\Re \left\{ 1/\alpha^{\rm co}_{\rm ee} \right\}=(\omega_0^2-\omega^2) /A$, where \mbox{$A=3000$ $\rm m^2 \. rad^2 / (s \. Ohm)$} and $\omega_0=2.81\.10^{10}$ $\rm rad / s$ have been chosen to correlate with the numerical results described in the next section.
It is seen  that the amplitude of the transmitted wave  is identically equal to unity at all frequencies, while its phase spans a full $2\pi$ range (the arctangent function in (\ref{eq:phase3}) varies over $\pi$, therefore, $\phi_{\rm t}$ varies over $2\pi$). Similar frequency dispersions were explored in \cite{kivshar}.
Since in our transmitarray all the unit cells should operate at the same frequency, the required phase variations can be achieved by adjusting the polarizability $\alpha^{\rm co}_{\rm ee}$  according to  (\ref{eq:phase3}). The simplest way to control the polarizability strength of the unit cell is to proportionally scale  all the sizes of its inclusions. As seen from Fig.~\ref{fig:fig3a}, at the resonance ($4.47$~GHz), the phase of transmission is $-\pi$. If we fix this frequency as the operational one, downscaling all the dimensions of the unit-cell inclusions will result in  a phase increase (from $-\pi$ towards 0)  of the transmitted wave at the operational frequency. Upscaling the inclusions, vice versa, will lead to a phase decrease (from $-\pi$ towards $-2\pi$).

%as we decrease the inclusions dimensions the resonance frequency becomes higher and the phase curve will be  shifted to the right so the phase at the operating frequency becomes higher and vise versa.   

It is simple to prove that a metasurface possessing only electric dipole response ($\widehat{\alpha}^{\rm co}_{\rm mm}=0$) cannot provide full phase variation of transmission. Indeed, in this case reflections from the metasurface inevitably appear $\mathbf{E}_{\rm r}\neq0$ and the phase of the transmitted wave 
\begin{equation}
\displaystyle
\phi_{\rm t}=\arctan \left(\frac{\omega \eta_0}{2S}\frac{1}{\Re \left\{
1/\alpha^{\rm co}_{\rm ee}  \right\}
 }\right) 
 \label{eq:phase4}
\end{equation}
spans only the $\pi$ range. Therefore, metasurfaces possessing solely electric dipole response (commonly called in the literature as single-layer frequency selective surfaces)  cannot have 100\% efficiency \cite{capasso,shalaev,grady}.

In summary, our analysis shows that broadband reflectionless uniaxial transmitarrays can be realized only with bianisotropic single-wire inclusions whose magnetoelectric coupling is compensated on the level of the unit cell. In this case, the polarization of the transmitted wave is the same as that of the incident one. Importantly, polarization plane rotation is impossible in such transmitarrays.

\section{Synthesis of broadband reflectionless transmitarrays}

Based on the preceding theoretical analysis, we  synthesize transmitarrays from chiral helical inclusions (see Fig.~\ref{fig:fig2b}), compensating chirality on the level of the unit cell. 
Alternatively, one could use inclusions with omega electromagnetic coupling. 
Without loss of generality, in this paper we design transmitarrays operating in microwaves on account of peculiarities of the inclusions fabrication. Arrays of
helical inclusions operating at infrared frequencies can be
manufactured based on fabrication technologies reported
in \cite{fabr1,fabr2}.
%\subsection{Choosing the unit cell geometry}

\begin{figure}[h]
\centering
 \subfloat[]{{\includegraphics[width=0.45\columnwidth]{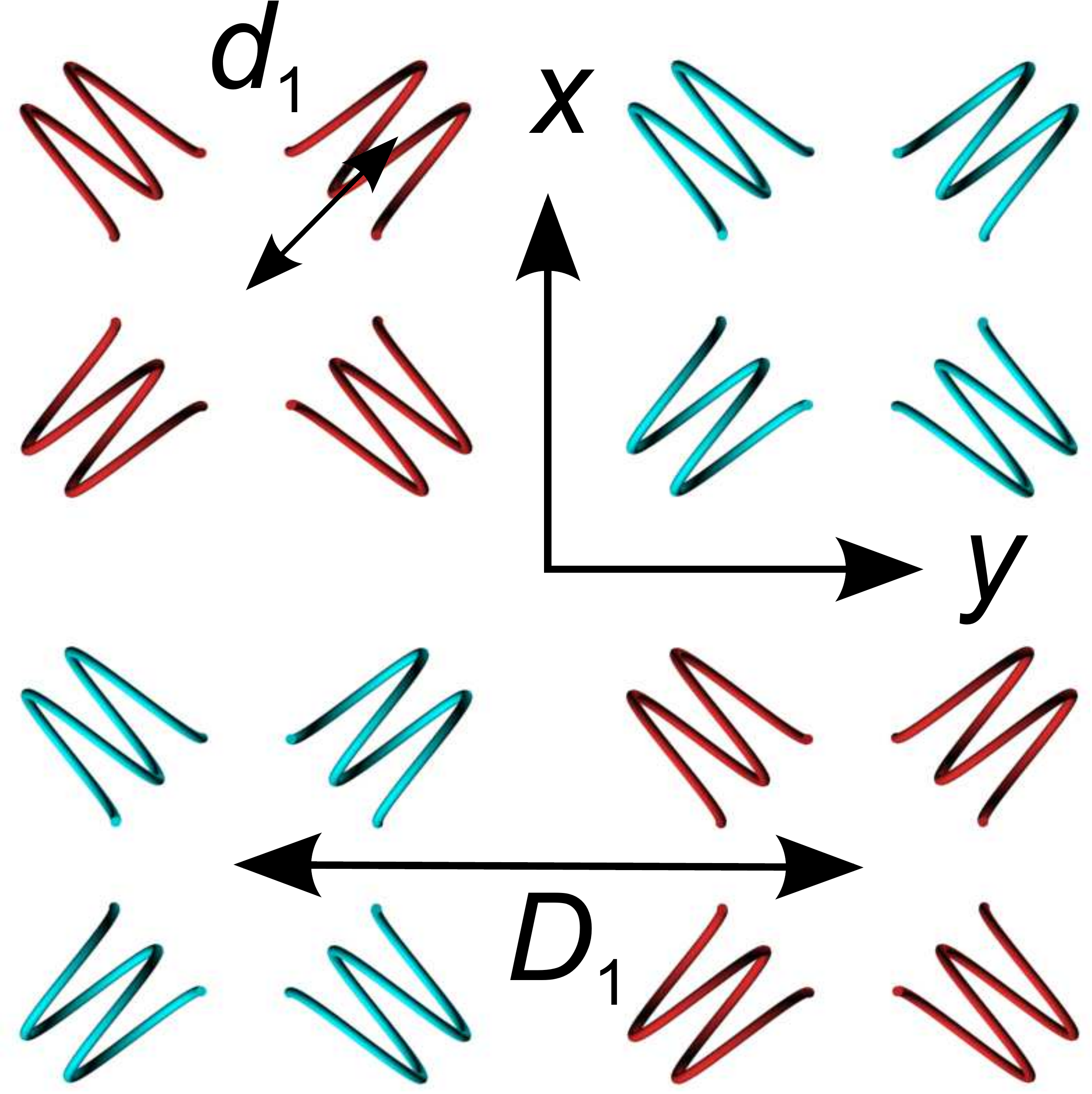} }
  \label{fig:fig4a}} 
 \quad
 \subfloat[]{{\includegraphics[width=0.45\columnwidth]{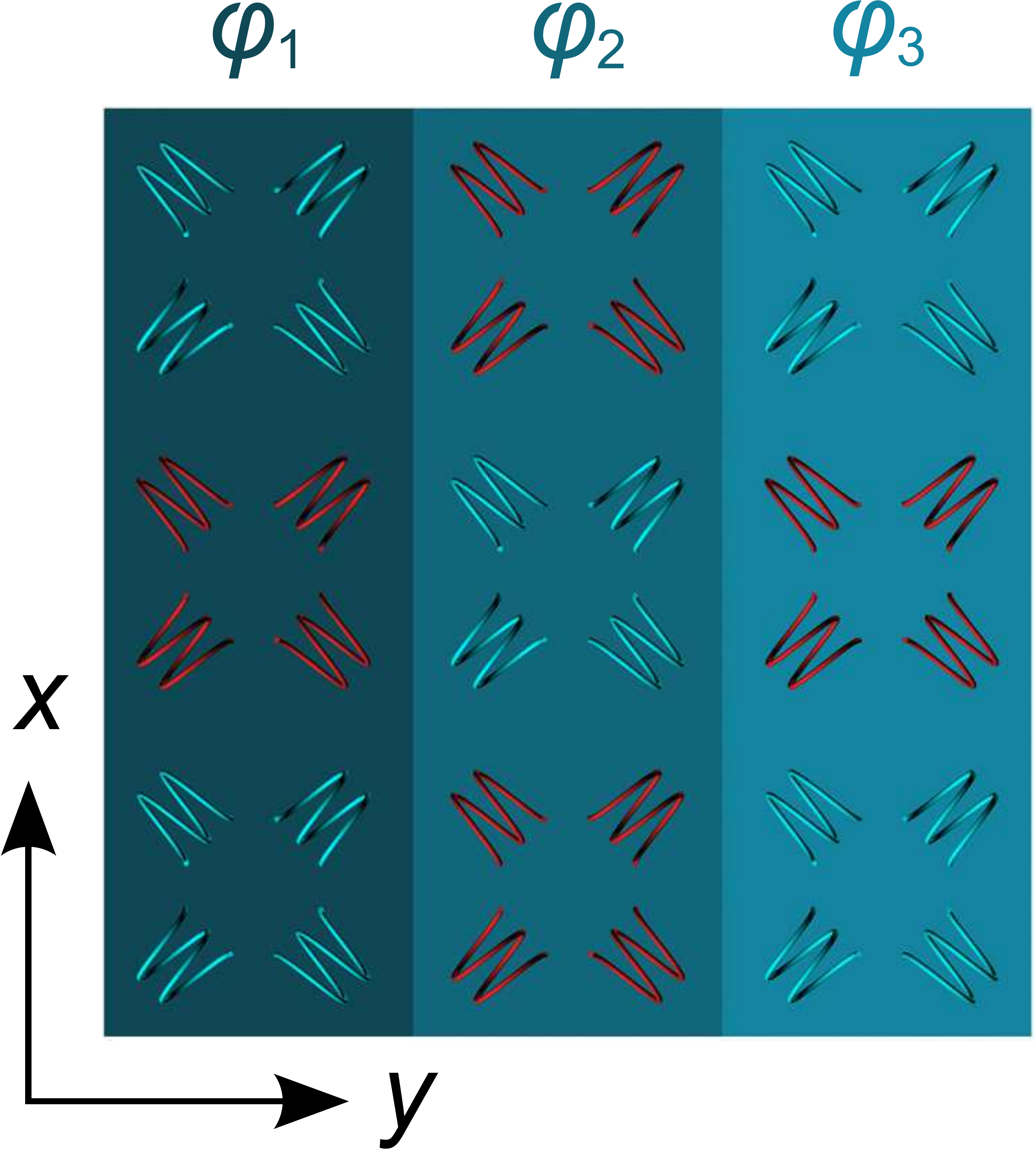} }
  \label{fig:fig4b}} 
 \caption{(a) Arrangement of the inclusions in the unit cell. Left- and right-handed inclusions are shown in red and blue, respectively. (b) Phase variations over the transmitarray. Different background colors denote blocks of helices with different phases $\phi_1$, $\phi_2$ and $\phi_3$.
 }
 \label{fig:fig4}
\end{figure}

First, it is important to design the unit-cell topology with suppressed chirality. To ensure uniaxial symmetry,  the unit cell should contain helices oriented in  two orthogonal directions in the metasurface plane.
We utilize the arrangement of helices   proposed in \cite{arrangement,PRX} and shown in Fig.~\ref{fig:fig4a}. 
The unit cell includes two blocks of left-handed and two blocks of right-handed helices. 
The sub-wavelength size of the inclusions ensures that the unit-cell size $2D_1$ is smaller than the operational wavelength. Therefore, the array of such unit cells can be modelled as sheets of homogeneous surface electric and magnetic currents, and the reflected and transmitted plane-wave fields are determined by  expressions (\ref{eq:q31}) and (\ref{eq:q32}). 

Figure~\ref{fig:fig3b} shows numerically calculated \cite{hfss} amplitude and  phase of transmission coefficients through an infinite periodic array of the unit cells shown in Fig.~\ref{fig:fig4a}. The unit-cell dimensions in this example were chosen as follows: $D_1=14.14$~mm, $d_1=5$~mm (the distance between the center of the block and the center of helices). The helices have the  pitch (the height of one turn) $l_{\rm ch}=1.38$~mm, and the radius of the turn $R_{\rm ch}=2.15$~mm. The radius of the inclusion wire is $r_0=0.33$~mm. 
As one can see from Fig.~\ref{fig:fig3b}, the transmittance  is more than 88\% at all frequencies, while the phase of transmission spans nearly full $2\pi$ range from 3.5~GHz to 5.5~GHz. In contrast to the theoretical results in Fig.~\ref{fig:fig3a}, in this case transmission is not unity at the resonance due to some dissipation of energy in copper helices.

As it was discussed in the previous section, the phase control of the transmitted waves can be accomplished by proportional scaling  the inclusions dimensions. 
The phase variation is engineered, for simplicity, only along one direction, along the $y$-axis.
In our design of transmitarrays with a non-uniform phase distribution we tune the phase individually for each block of helices (not the entire unit cell) to ensure smoother phase gradient over the transmitarray plane (see Fig.~\ref{fig:fig4b}). Although in this case chirality of adjacent in the $y$-direction blocks is not completely compensated (because the helices in the blocks have slightly different sizes and polarizability amplitudes),  overall, the chirality effect is nearly suppressed due to a great number of different unit cells. 

Based on the preceding theoretical analysis, we design two  transmitarrrays with different functionalities in order to demonstrate the potential of the approach. These examples show how to manipulate the direction of wave propagation as well as the wavefront shape.

\subsection{Manipulating the direction of wave propagation}
In this example we synthesize a transmitarray that refracts normally incident waves (along the $-z$-direction) at an angle $45\degree$ in the $yz$-plane. 
To achieve the effect of anomalous refraction, we need to tune the inclusions dimensions in every block so that there is a linear phase gradient of transmission along  the $y$-direction of the array. Thus, from the phased arrays theory, the array should be periodical along the $y$-direction with the period $ d=\lambda /\sin 45\degree=98.2$~mm, where $\lambda$ is the  wavelength at the operational frequency 4.32~GHz. The phase of transmission changes from 0 to $2\pi$ along one period $d$. 
The periodicity of the array in the $x$ direction is $2D_1$ (the period of the unit cell), since along this direction there is no phase variation. 
In order to ensure smooth phase variations, we place the maximal number of inclusions blocks with prescribed phases along the period $d$. Based on the dimensions of the helices (about $\lambda/12$), we form the period of six blocks of helices, i.e. $D_1=d/6=16.4$~mm. In this example the spacing between the helices in the blocks $d_1=5.75$~mm. The dimensions of the helices in each block are listed in Table~\ref{tabl:table2} in Appendix~\ref{app:1}.

\begin{figure}[h]
\centering
 \subfloat[]{{\includegraphics[width=0.70\columnwidth]{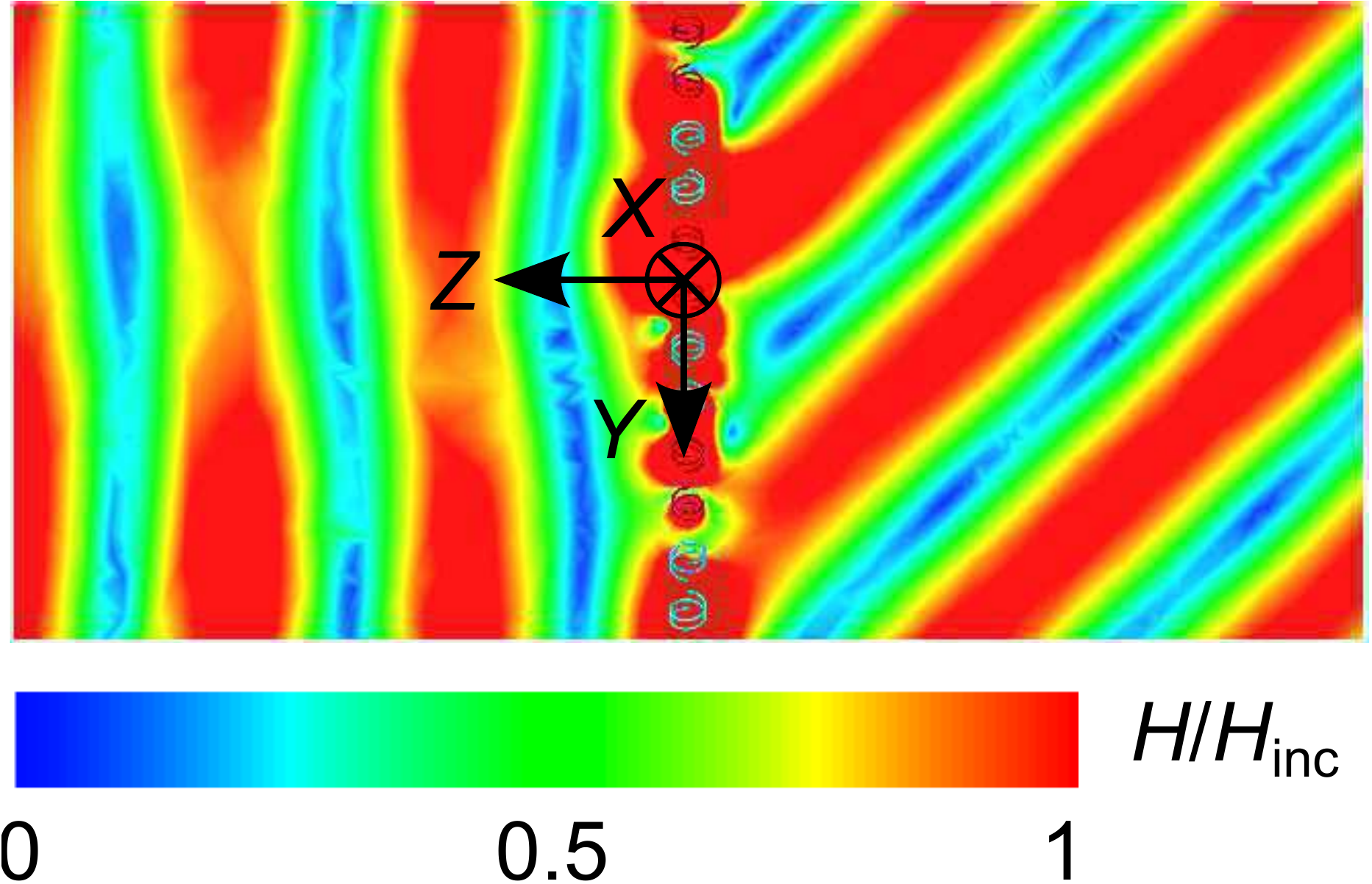} }\label{fig:fig5a}}%
 \qquad
 \subfloat[]{{\includegraphics[width=0.80\columnwidth]{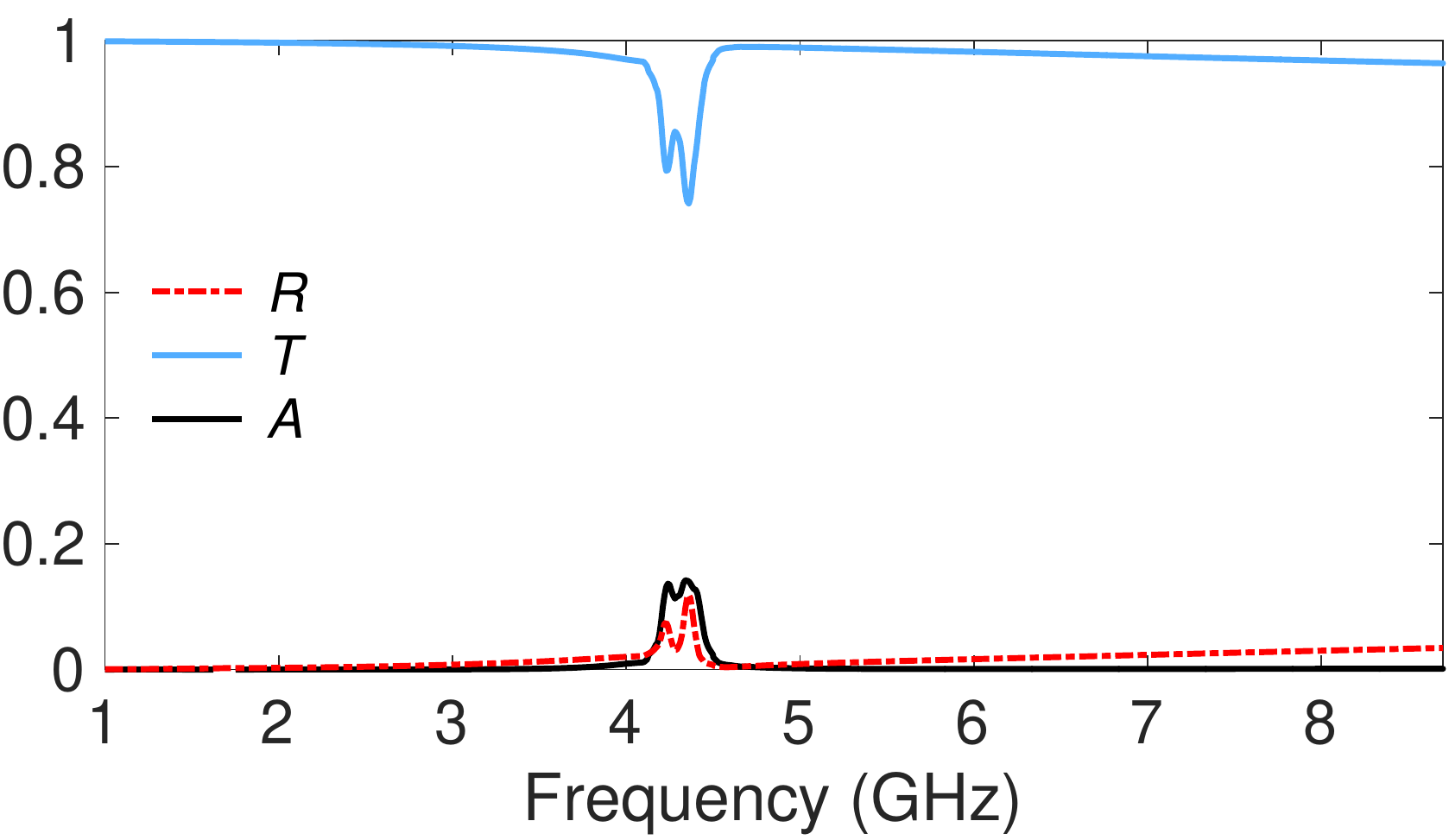} }\label{fig:fig5b}}%
 \caption{(a) Magnetic field distribution normalized to the magnetic field of the incident wave. The incident wave propagates along the $-z$-direction with the electric field along the $x$-axis. (b) Reflectance $R$, transmittance $T$ and absorbance $A$ versus frequency.}%
 \label{fig:fig5}%
\end{figure}

The simulated results for the designed transmitarray are shown in Fig.~\ref{fig:fig5a}. 
Indeed, the structure refracts the incident wave at $45\degree$ from the normal. At the operating frequency 4.32~GHz (see Fig.~\ref{fig:fig5b}) the transmittance from the structure reaches 83\%. Non-zero reflection of 5\%  and absorption of 12\% in the transmitarray result from the non-ideal impedance equalization \cite{passive}. 
Remarkably,  the transmitarray passes through more than 95\% of the incident power (without its modification) beyond its operational band from 4.13 to 4.47~GHz (see Fig.~\ref{fig:fig5b}). At very high frequencies some parasitic reflections from the transmitarray appear. They are caused by the higher-order resonances in the double-turn helices of the transmitarray and occur near the triple operating frequency at 13.2~GHz \cite{PRX}. At very low frequencies, the transmitarray inclusions are not excited by incident waves and, therefore, are nearly fully transparent.

%
%Using the chosen unit cell we build a finite array of only one strip by alternating right and left handed unit cells to compensate chirality. PEC boundary conditions will be applied on the top and bottom of the array in the numerical simulations as well as in the experiment by placing the array in a parallel plate wave guide. According to the image theory this will represent an infinite periodic array in the vertical x-direction. From all other sides we put PML as it will be finite horizontally. The PML will be represented by absorbers surrounding all sides of the parallel plate wave guide in the experiment phase. 
\subsection{Wavefront shaping}
In order to demonstrate the ability of wavefront shaping, we design a transmitarray that focuses normally incident plane waves in a line parallel to the $x$-axis. Due to reciprocity, the metasurface illuminated by a line source from the focal point transmits a collimated beam. Such lens performance requires that the phase gradient of the transmitarray has a parabolic profile.
The designed focal distance of the lens  is just a fraction of the operational wavelength $f=0.64\lambda$. Such a short focal distance is provided by the sub-wavelength sizes of the helices. The dimensions of the blocks of helices in this example are as follows: $D_1=14.14$~mm and $d_1=5$~mm.
The lens is infinite along the $x$-axis with the periodicity equal to the size of one unit cell $2D_1$. Along the $y$-direction the lens is $410.1$~mm long and contains 29 blocks of helices. The parabolic phase gradient dictated by 
\begin{equation}
\displaystyle
\phi_{\rm t}(y)= \phi_{\rm t}(0)+\frac{2\pi}{\lambda} \sqrt{y^2+f^2}
 \label{eq:phase55}
\end{equation}
is achieved due to precise tuning of the inclusions dimensions in each block (described in Table~\ref{tabl:table3} in Appendix \ref{app:1}). Here, $y$ is the coordinate, $\phi_{\rm t}(0)=-\pi$ is the phase of transmission in the center of the transmitarray (chosen arbitrarily)  and $\lambda$ is the wavelength at 4~GHz. 

\begin{figure}[h]
	\centering
	\hbox{\hspace{1cm}\includegraphics[width=0.7\columnwidth]{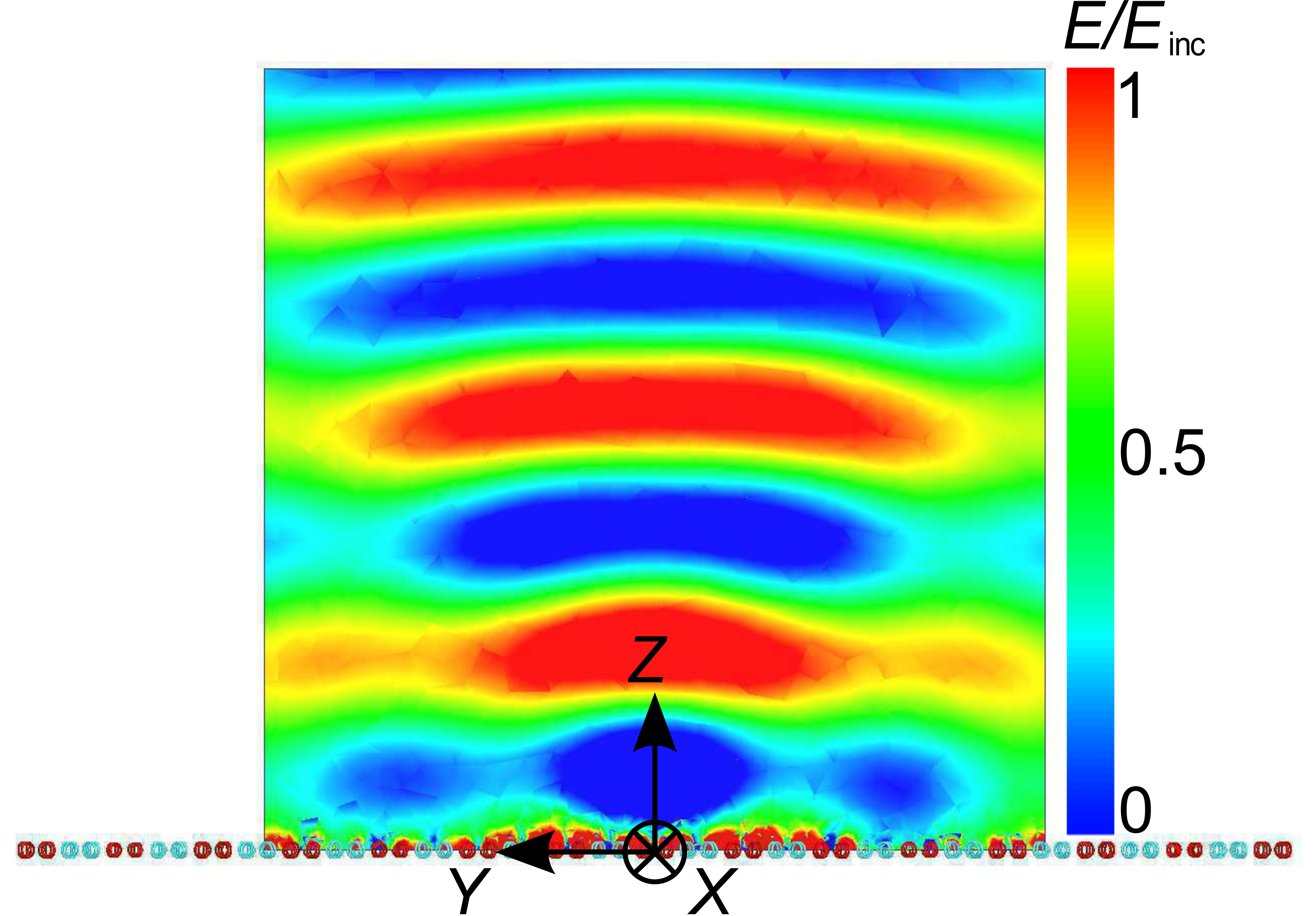}}
	\caption{Simulated electric field distribution of the wave transmitted through the lens. The source of incident cylindrical waves is located in the focal point at $z=-f$. The lens center is at the origin of the coordinate system. }
	\label{fig:fig6}
\end{figure}

To test the performance of the designed lens, we illuminated it by a source of cylindrical waves located at the focal distance from the lens. The simulation results at the operating frequency $3.9$~GHz (the actual frequency was shifted from the designed one) are presented in Fig.~\ref{fig:fig6}.
As expected, the lens transforms the cylindrical wavefront of the incident wave into a planar one. 

\begin{figure}[h]
\centering
 \subfloat[]{{\includegraphics[height=3cm, width=0.47\columnwidth]{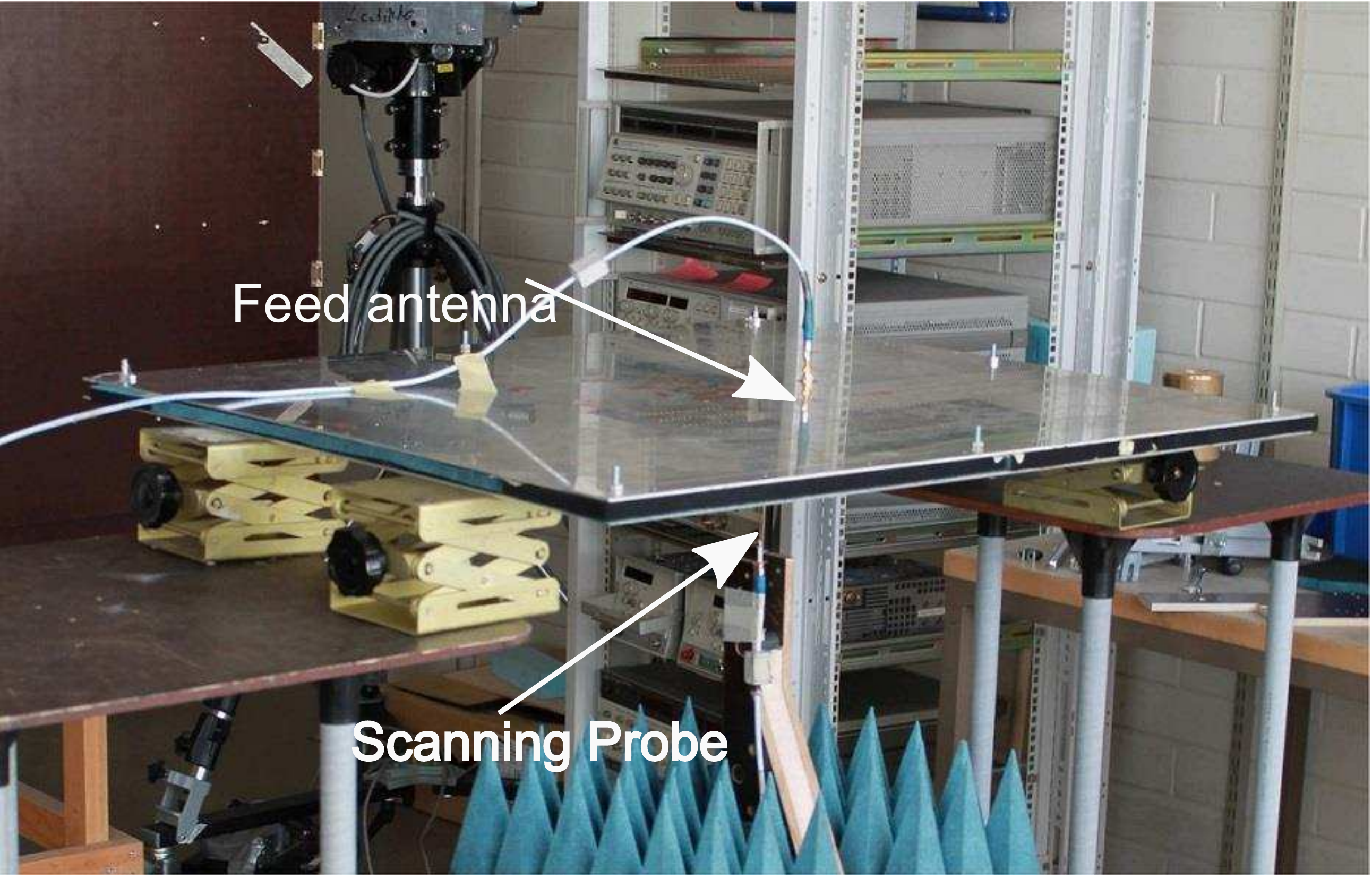} }\label{fig:fig7a}}
 \hspace{1mm}
 \subfloat[]{{\includegraphics[height=3cm, width=0.47\columnwidth]{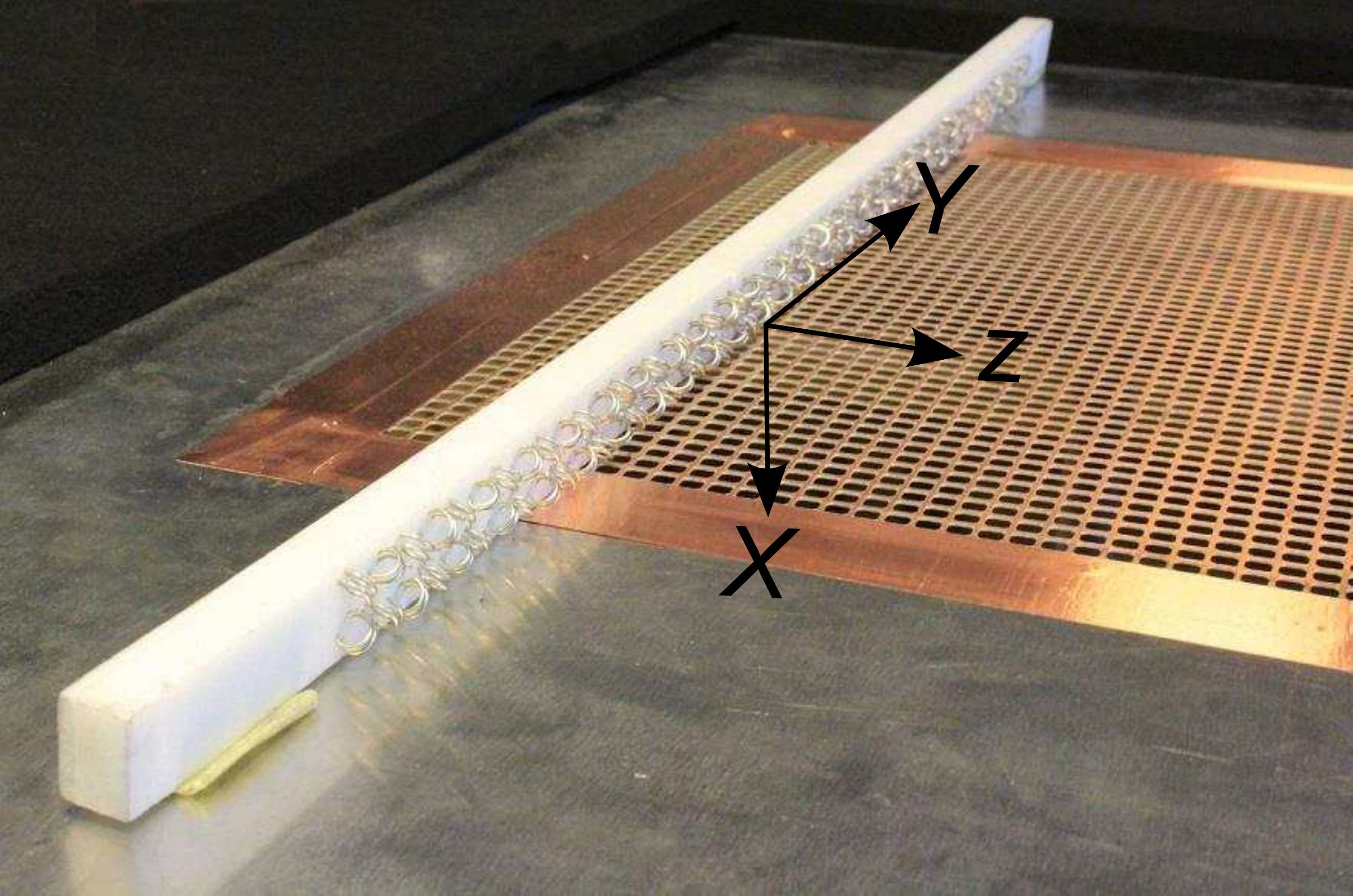} }\label{fig:fig7b}}%
 \caption{Experimental realization. (a) Experimental setup with the fabricated lens placed inside a parallel-plate waveguide. (b) Fabricated lens consisted of 29 blocks of helices providing parabolic phase variations along the $y$-axis. 
}%
 \label{fig:fig7}%
\end{figure}

Next, experimental testing of the designed lens was conducted in a parallel-plate waveguide (Fig.~\ref{fig:fig7a}).
According to the image theory, images of chiral inclusions placed between the plates of the waveguide represent equivalent chiral inclusions with the opposite handedness. Therefore, it is enough to place only one row of blocks (one half of each unit cell) inside the waveguide (see Fig.~\ref{fig:fig7b}). Effectively, it  emulates full unit cells (Fig.~\ref{fig:fig4a}) periodically repeated along the $x$-direction. The helical inclusions were fabricated  with precision 0.01~mm and embedded in Rohacell-51HF material with $\epsilon_{\rm r}=1.065$ and $\tan\delta=0.0008$ for mechanical support.
The transmitarray was excited by a monopole antenna oriented along the $x$-axis and placed in the focal point at $z=-49$~mm. 

\begin{figure}[h]
\centering
 \subfloat[]{{\includegraphics[width=0.8\columnwidth]{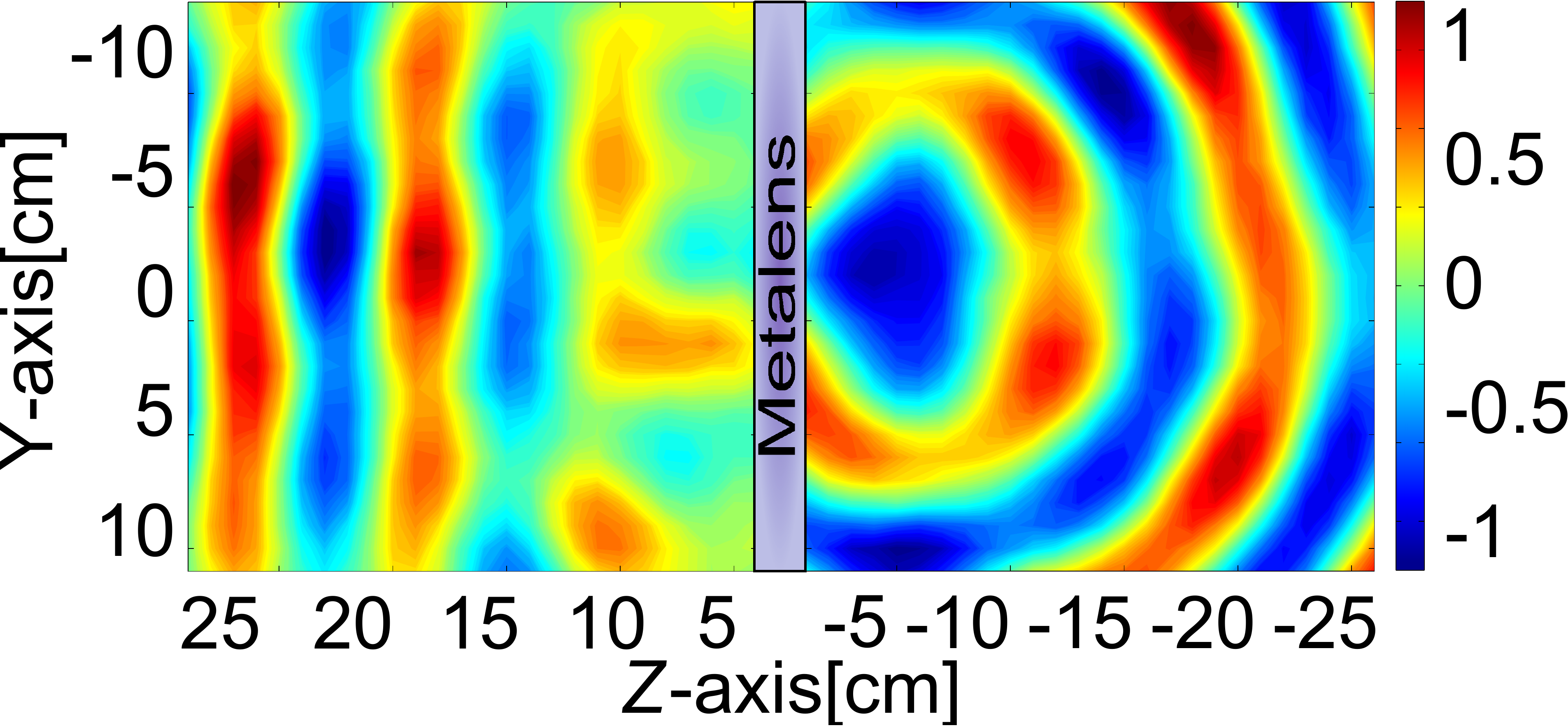} }\label{fig:fig8a}}
 \\
 \subfloat[]{{\includegraphics[width=0.48\columnwidth]{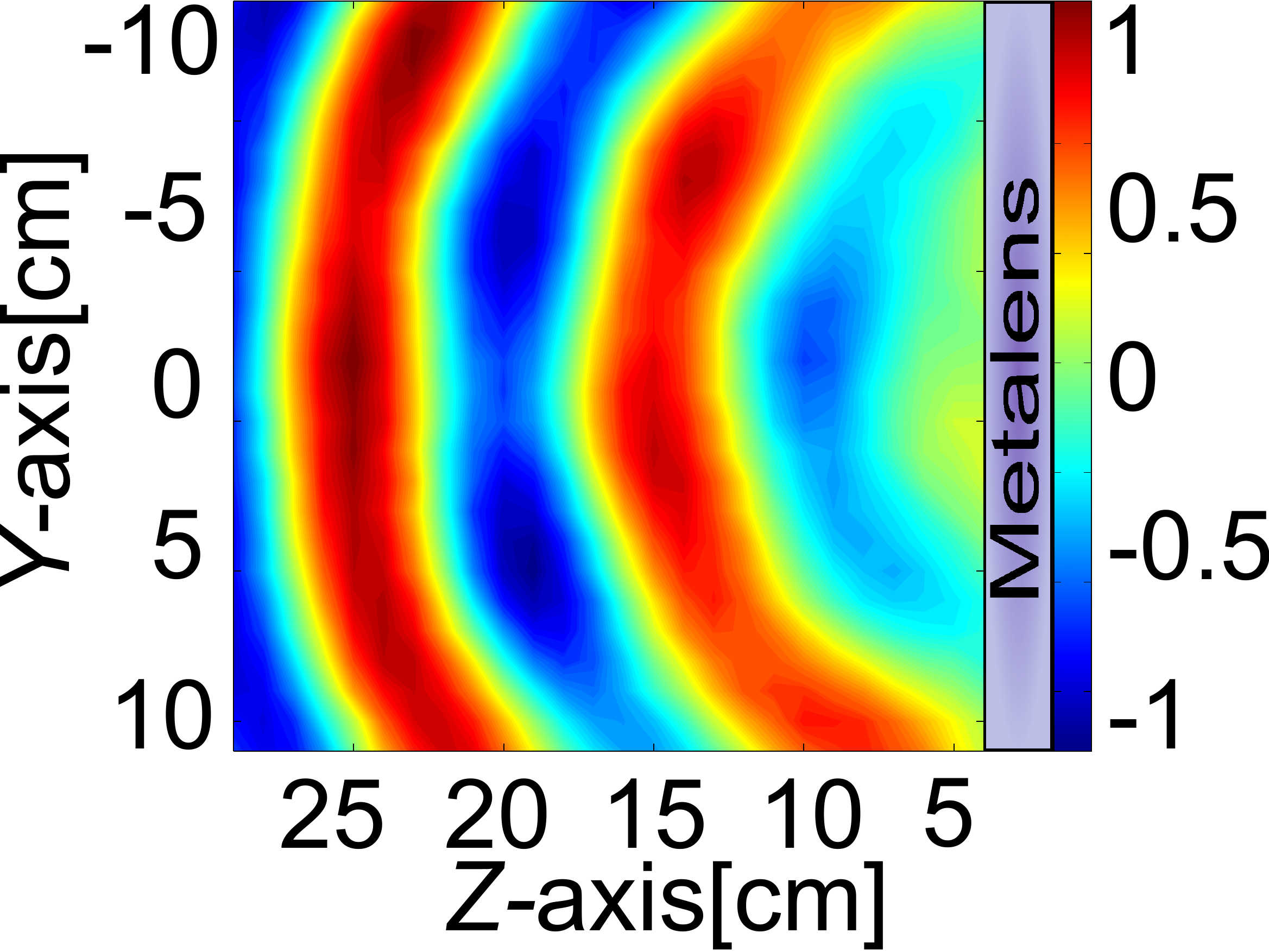} }\label{fig:fig8b}}
\hspace{0mm}
 \subfloat[]{{\includegraphics[width=0.48\columnwidth]{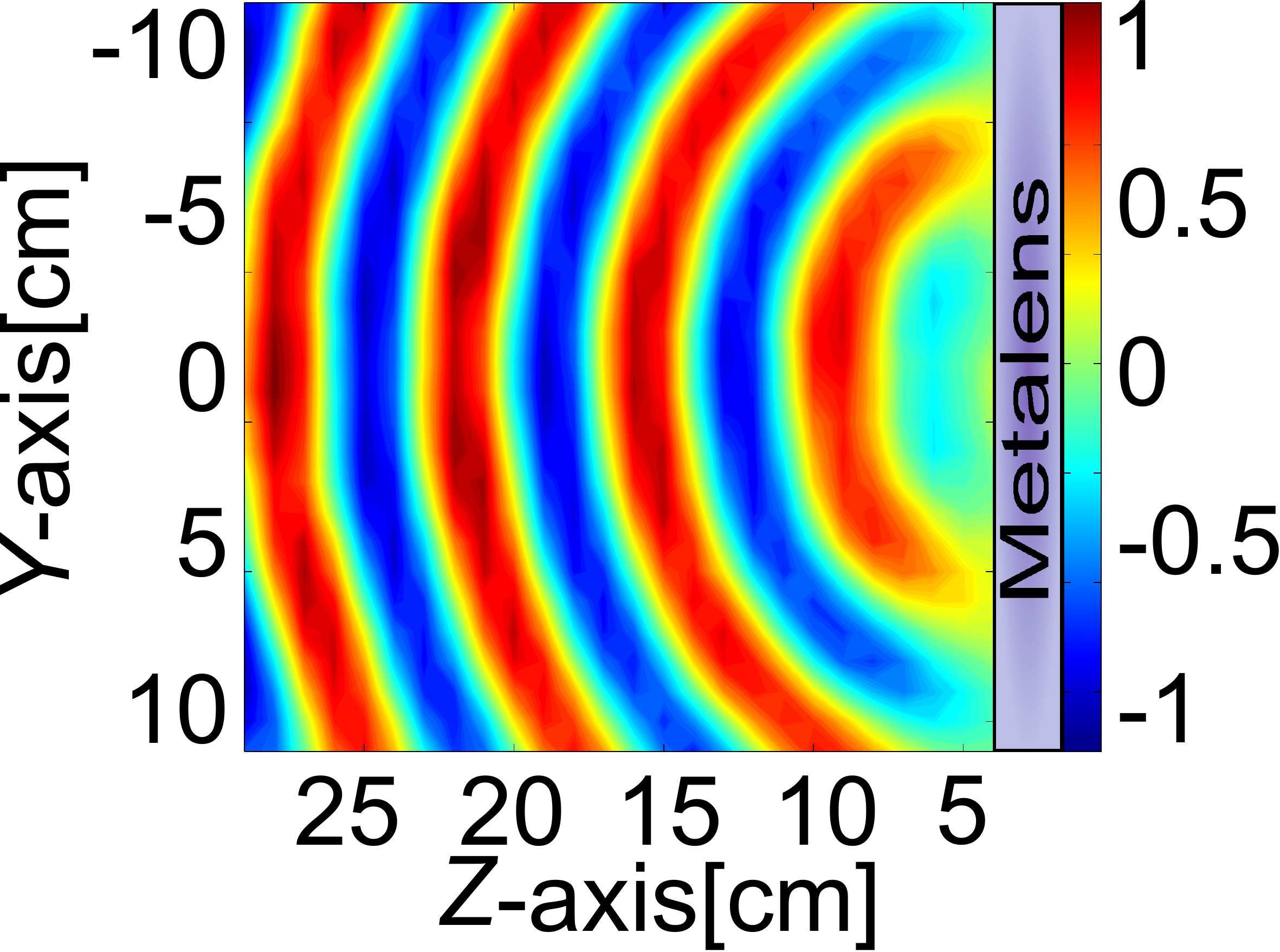} }\label{fig:fig8c}}%
 \caption{Distribution of the measured distribution of the $x$-component of the electric field inside the waveguide:  (a) at the resonance frequency 3.86~GHz, (b) at frequency 3~GHz and (c) at frequency 5~GHz. The feed antenna is located at point $z=-49$~mm. The lens location is shown by the blue box. 
 }%
 \label{fig:fig8}%
\end{figure}

The bottom plate of the waveguide incorporates a copper mesh with the period of 5~mm (see Fig.~\ref{fig:fig7b}). Due to the deeply sub-wavelength periodicity, the mesh practically does not disturb the fields inside the waveguide. 
On the other hand, outside of the waveguide there are decaying fields in the near proximity of the mesh. The electric field distribution inside the waveguide can be analysed  through these near fields  measured  by a small probe antenna (Fig.~\ref{fig:fig7a}). 
More detailed information about the measurement set-up can be found in \cite{joni,FuncMetamirror}.
The measured electric field distribution inside the waveguide at the resonance frequency  3.86~GHz is shown in Fig.~\ref{fig:fig8a}.  
One can see that the fabricated lens in fact transforms the cylindrical wavefront of the incident wave into a planar one. According to Fig.~\ref{fig:fig8b} and Fig.~\ref{fig:fig8c}, as expected, the lens does not interact with  the incident waves beyond the operational band. Incident waves pass through the structure without attenuation and wavefront transformations. 
This experimental result confirms our theoretical findings.

\section{Multifunctional cascaded metasurfaces}
\begin{figure}[h]
\centering
		\subfloat[]{{\includegraphics[width=0.48\columnwidth]{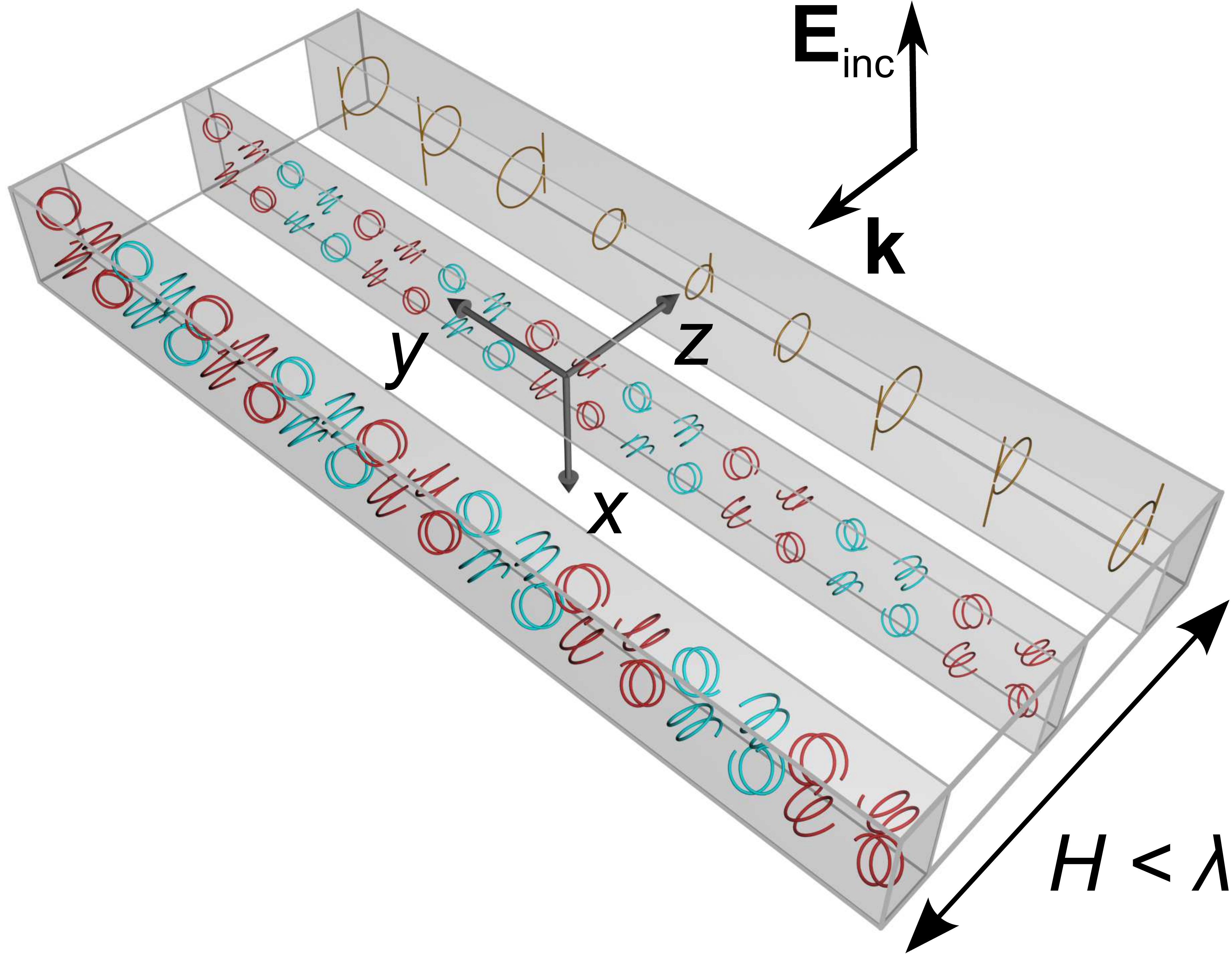} }\label{fig:fig9a}}%
\hspace{0mm}
		\subfloat[]{{\includegraphics[width=0.48\columnwidth]{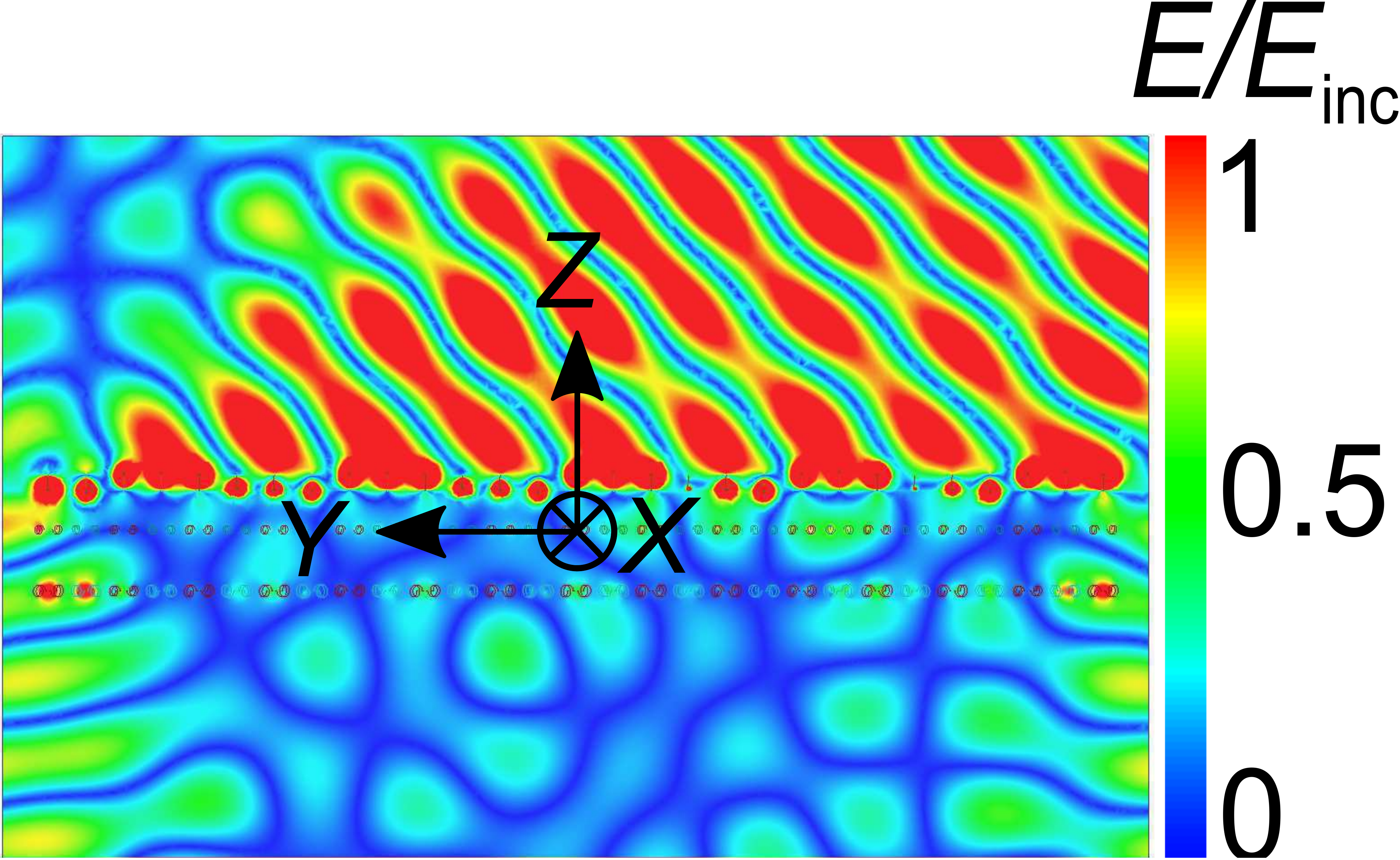} }\label{fig:fig9b}}%
\hspace{0mm}
		\subfloat[]{{\includegraphics[width=0.48\columnwidth]{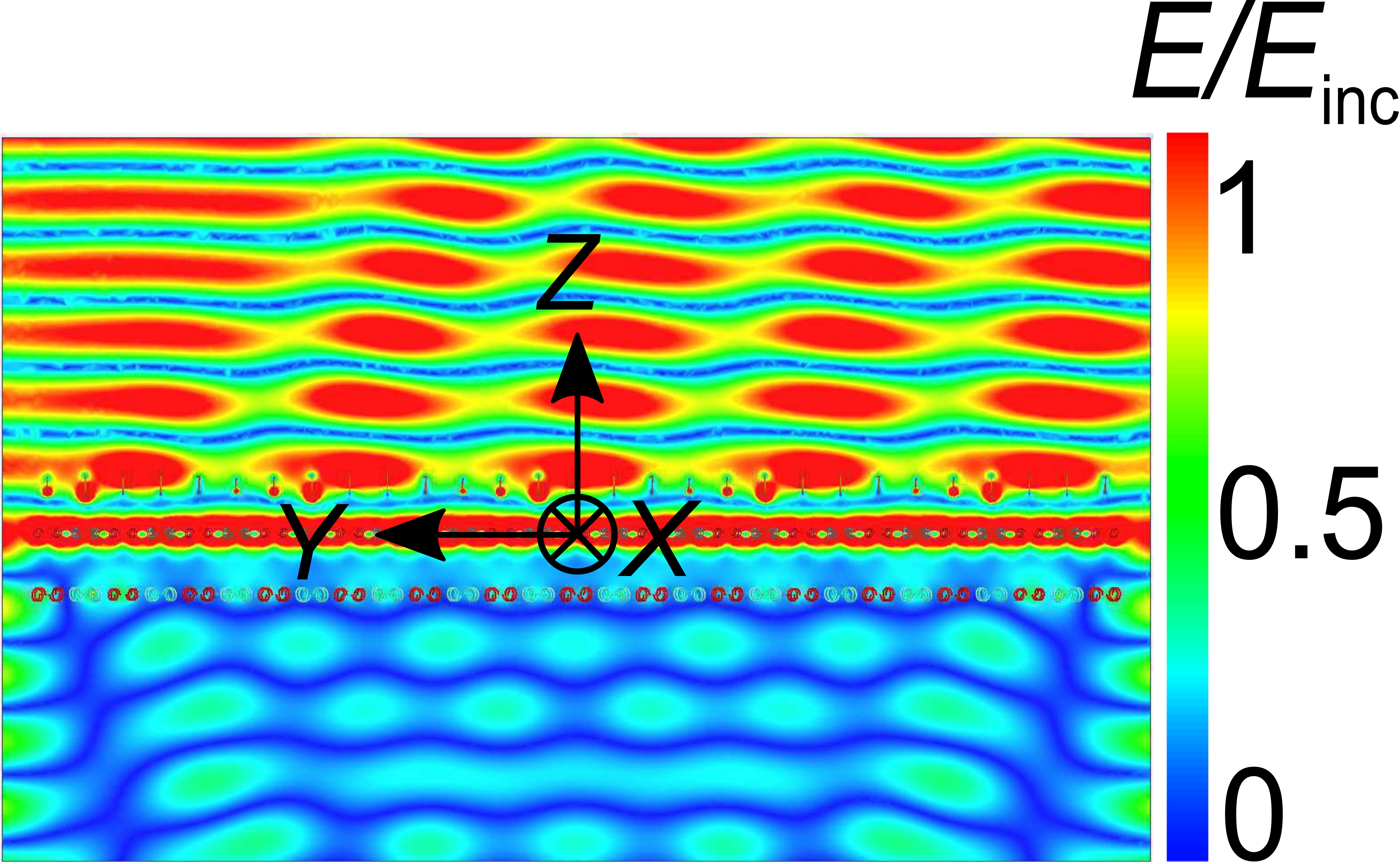} }\label{fig:fig9c}}
\hspace{0mm}
		\subfloat[]{{\includegraphics[width=0.48\columnwidth]{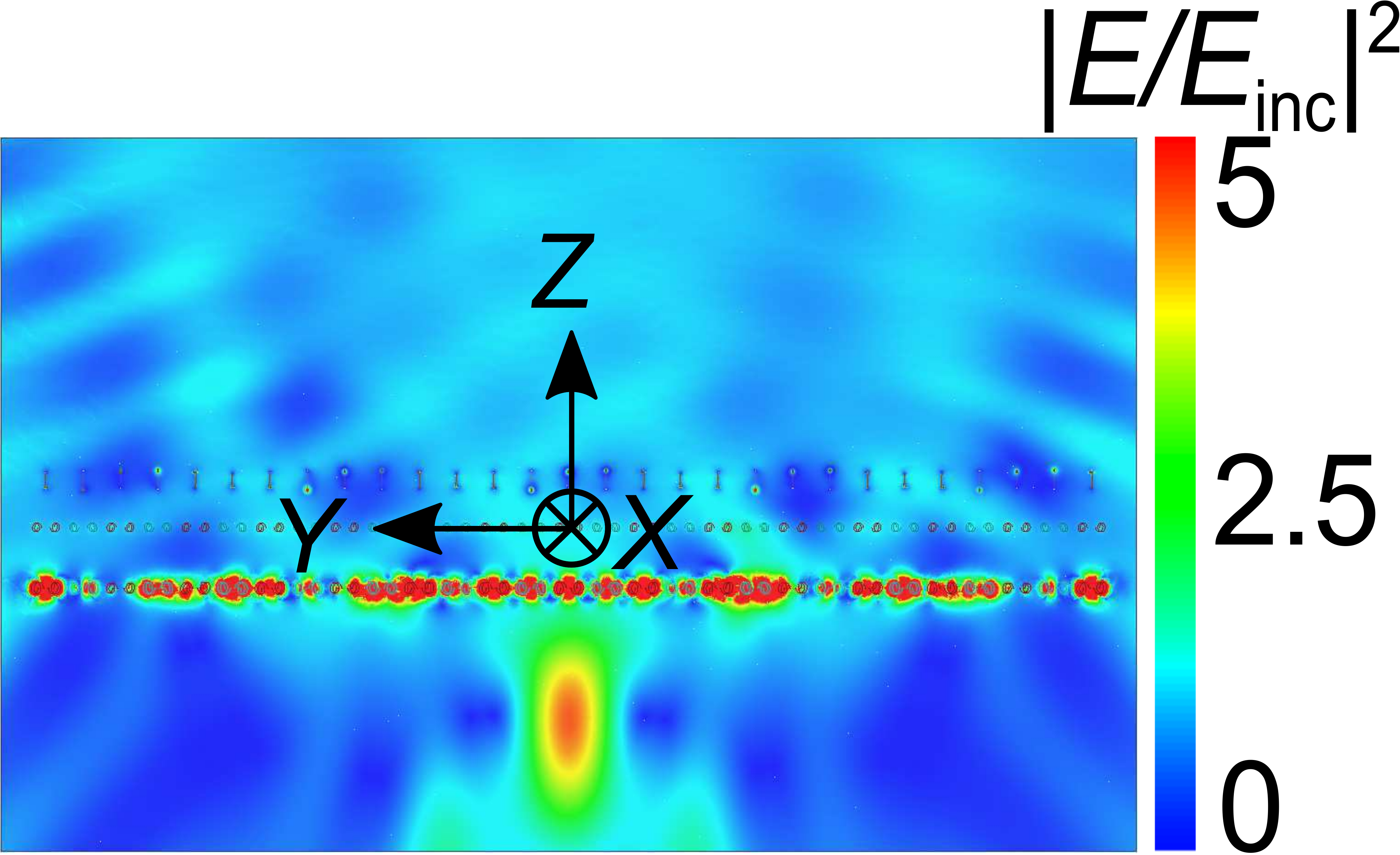} }\label{fig:fig9d}}
		\caption{(a) Cascade of three metasurfaces. Only 11 blocks are shown for clarity. (b) Simulated electric field distribution at 5~GHz. (c) Simulated electric field distribution at 6~GHz. (d) Simulated power distribution at 3.9~GHz. }
		\label{fig:fig9}
\end{figure}

In this section we explore the possibility for integration of the designed transmitarrays in a cascade of metasurfaces. To highlight the three basic functionalities for wave control, such as manipulation of reflection, transmission and absorption properties, we design and test numerically a composite layer consisting of three cascaded metasurfaces with the corresponding properties (see Fig.~\ref{fig:fig9a}). The incident wave illuminates the cascade normally from the $+z$-direction. The first metasurface illuminated by the incident wave is a so-called metamirror proposed in \cite{FuncMetamirror}. It nearly fully reflects normally incident waves at 5~GHz at an angle $45\degree$ from the normal. The second metasurface was designed to totally absorb incident radiation at 6~GHz. It represents a composite of double-turn helices similar to that described in \cite{PRX} but tuned to operate at another frequency. All the helices in the composite have the same dimensions: the helix pitch is 
$l_{\rm ch}=1.11$~mm, the helix radius $R_{\rm ch}=1.71$~mm and the wire radius $r_0=0.1$~mm. The helices are made of lossy nichrome $\rm NiCr60/15$ with the conductivity about $10^6$~S/m. The third cascaded metasurface is the lens designed in the present work and operating at 3.9~GHz. All the metasurfaces consist of 29 blocks and have the same spacing 14.14~mm between the adjacent blocks.

The second (middle) metasurface is located at  the origin of the coordinate system, while the first and the third structures are positioned at $z=18$~mm and $z=-23$~mm, respectively. Such spacing  ensures that the metasurfaces are located from one another at a distance not less than $\lambda/3$ at their operational frequencies to prevent strong near-field interactions. The overall thickness of the three-layer structure is $H=48$~mm, which  does not exceed one wavelength at 6~GHz. 

The performance of the metasurface cascade at the three operating frequencies is shown in Figs.~\ref{fig:fig9b}, \ref{fig:fig9c} and \ref{fig:fig9d}. 
At 5~GHz, incident waves are nearly totally reflected by the first metasurface at the angle $45\degree$ from the $z$-axis (Fig.~\ref{fig:fig9b}). At 6~GHz, the first metasurface becomes ``invisible'' for incident waves, and nearly all the power is absorbed by the second metasurface (Fig.~\ref{fig:fig9c}). Finally, incident waves at 3.9~GHz pass through the first two metasurfaces and are focused by the third metasurface (Fig.~\ref{fig:fig9d}) nearly at the designed focal distance $f=0.64\lambda$. This sub-wavelength three-layer composite is equivalent to the structure shown in Fig.~\ref{fig:fig1c}.

The data for reflection, transmission and absorption properties of the three-layer structure is summarized in Table~\ref{tabl:table1}. Since the structure has a finite size, reflectance and transmittance were introduced as ratios of reflected and transmitted powers to the power incident  through the cross section of the metasurfaces. 
\begin{table}[h]
\centering
\scriptsize
%\footnotesize
\begin{tabular}{|>{\centering}p{0.2\columnwidth}|>{\centering}p{0.2\columnwidth}>{\centering}p{0.2\columnwidth}p{0.2\columnwidth}<{\centering}|} 
\hline
& & &   \\[-2.0mm]
Frequency, GHz &  Transmittance $T$, \%& Reflectance $R$, \% &  Absorbance $A$, \%  \\ \hline\hline
& & & \\[-2.0mm]
2.0     & 99.6    & 0.2   & 0.2    \\
3.0     & 96.8    & 3.0   & 0.2    \\
3.9     & 59.2  & 26.8 & 14.0  \\
5.0     & 8.0   & 85.8 & 8.3   \\
6.0     & 7.9   & 5.1  & 86.5  \\
7.0     & 84.3    & 12.5    & 3.2    \\
8.0     & 84.6    & 10.7  & 4.7    \\
\hline
\end{tabular}
\caption{Numerically calculated characteristics of the metasurface cascade.}
\label{tabl:table1}
\end{table}
As one can see from Table~\ref{tabl:table1}, while reflection and absorption levels at 5 and 6~GHz, respectively, are high (more than 85\%), transmission level  at 3.9~GHz is moderate (about 60\%). This can be explained by two factors. First, there are some  diffraction effects at the edges of the three finite-size metasurfaces.
Second, the spectrum separation between the metasurfaces operating at 3.9 and 5~GHz is not high enough. The metamirror still reflects a small part of the incident energy at 3.9~GHz. It is seen from Table~\ref{tabl:table1}, far from the operating frequencies transmission of incident waves through  the metasurface cascade exceeds 84\%.

\section{Conclusion}

In this paper, we have proposed a new type of transmitarrays that allow full wave control (with the efficiency more than 80\%) and are transparent beyond the operating frequency range. 
Due to the frequency-selective response of the designed transmitarrays, they can be easily integrated in existing and new complexes of antennas and filters. 
In this paper, we have also proposed  an approach for designing multifunctional cascades of metasurfaces. Depending on the frequency of incident radiation, such a cascade possesses different responses at different frequencies that can be carefully adjusted. To test the approach, we have designed a cascade of three metasurfaces that performs three different functions for wave control at different frequencies. Despite the multifunctional response, the thickness of the designed structure is smaller than the operational wavelength.  

Unique functionalities of the cascaded metasurfaces could be useful in a variety of new applications. Importantly, going to the limiting case of cascading metasurfaces, one can design a \emph{single} metasheet that incorporates different kinds of inclusions performing a  multifunctional response. Moreover, our approach of cascaded metasurfaces can be also extended to volumetric metamaterials. 

The main challenge for implementing the designed structures are fabrication issues. However, as we hope to show in our future work, the three-dimensional shape of the helical inclusions can be modified into an appropriate fabrication-friendly printed topology.

%By moving from conventional antenna systems to metasurfaces better performance, easier fabrication process, and easier integrability were achieved. Possible cases were studied theoretically in order to get a thin layer transmit array that is able to fully control the transmitted wave phase. We started assuming the array to be reciprocal and the inclusions to be wire and doesn't produce omega coupling. By allowing only Co-transmission it was proved that that will lead to using anistropic inclusions which will not allow matching over wide range of frequencies. By allowing only cross-transmission it was proved that we can get full phase control with chiral inclusions but they shouldn't be wire inclusions. Then we studied the case where we allow both Co \& Cross transmission which proved the possibility for full phase control but again for all general chiral inclusions except for chiral wires. Then we checked the case where we compensate chirality and it proved full phase control for any chiral inclusions used. Using these results the unit cell was chosen and a refraction and focusing arrays were designed and simulated. The refracting array showed around 4\% reflection at the operating frequency and less than 10\% near it and almost transparent otherwise. The focusing array was tested experimentally and showed close results to the simulations. Finally a 3 layer structure that absorb, reflect, and focus depending on the operating frequency and transparent otherwise was simulated and proved that our theory and design are realizable. 

\appendices
\section{}
\label{app:1}
%The $45\degree$ Refracting transmit array used in simulations consists of 6 unit cells each of 4 identical inclusions made of ``0.33mm raduis'' copper wire and oriented as shown in figure~\ref{fig:Arraystructure} where the red helicies are left handed and the cyan ones are right handed. The Pitch angle is $6.35\degree$ and unit cell radius is 5.75mm. The inclusions dimentions and their locations with respect to their position within the period are presented in table~\ref{atab:1}.\\
\begin{table}[h]
\scriptsize
%\footnotesize
\begin{tabular}{|>{\centering}p{0.2\linewidth}>{\centering}p{0.13\linewidth}>{\centering}p{0.13\linewidth}>{\centering}p{0.13\linewidth}p{0.15\linewidth}<{\centering}|} 
\hline
& & & &  \\[-2.0mm]
Location of the block along the $y$-axis within the period $d$ &  Handedness of the helices in the block & Loop radius  of the helices $R_{\rm ch}$, mm &  Pitch of the helices $l_{\rm ch}$, mm &  Phase of   waves transmitted through the block \\ \hline\hline
& & & &\\[-2.0mm]
$-5d/12$  & Left  & 2.34 & 1.63 & $-\pi/3$ \\
$-d/4$  & Right  & 2.37 & 1.66 & $-2\pi/3$ \\
$-d/12$  & Left  & 2.39 & 1.67 & $-\pi$ \\
$d/12$  & Right  & 2.41 & 1.68 & $-4\pi/3$ \\
$d/4$  & Left  & 2.44 & 1.71 & $-5\pi/3$ \\
$5d/12$  & Right  & 2.70 & 1.88 & $-2\pi$ \\ 
\hline
\end{tabular}
\caption{Dimensions of the helices in each of 6 blocks constituting period $d$. The radius of the wire $r_0=0.2$~mm.}
\label{tabl:table2}
\end{table}
%The Metalens used in simulations and experiment consists of 29 unit cells with the same structure and wire radius as the $45\degree$ Refracting transmit array. The pitch angle is $5.6\degree$ and unit cell radius is 5mm. The inclusions dimentions starting from the middle unit cell going through the array in +ve Y direction are presented in table~\ref{atab:2}, the same dimensions are repeated in -ve Y direction due to symmetry.\\
\begin{table}[h]
\scriptsize
%\footnotesize
\begin{tabular}{|>{\centering}p{0.2\linewidth}>{\centering}p{0.13\linewidth}>{\centering}p{0.13\linewidth}>{\centering}p{0.13\linewidth}p{0.15\linewidth}<{\centering}|} 
\hline
& & & &  \\[-2.0mm]
Distance from the center of the block to the center of the lens, mm &  Handedness of the helices in the block & Loop radius  of the helices $R_{\rm ch}$, mm &  Pitch of the helices $l_{\rm ch}$, mm &  Phase of   waves transmitted through the block, $^\circ$ \\ \hline\hline
& & & &\\[-2.0mm]
0.0   & Left   & 2.50 & 1.54  & 50 \\
14.1   & Right  & 2.48 & 1.53 & 60 \\
28.3   & Left   & 2.45 & 1.51 & 87 \\
42.4   & Right  & 2.43 & 1.50 & 127 \\
56.6   & Left   & 2.41 & 1.48 & 176 \\
70.7   & Right  & 2.38 & 1.46 & 230 \\
84.8   & Left   & 2.26 & 1.39 & 288 \\
99.0   & Right  & 2.59 & 1.60 & 348 \\
113.1  & Left   & 2.47 & 1.52 & 410 \\
127.3  & Right  & 2.43 & 1.50 & 473 \\
141.4  & Left   & 2.41 & 1.48 & 537 \\
155.5  & Right  & 2.36 & 1.46 & 601 \\
169.7  & Left   & 2.13 & 1.31 & 666 \\
183.8  & Right  & 2.52 & 1.55 & 732 \\
198.0  & Left   & 2.45 & 1.51 & 798 \\
\hline
\end{tabular}
\caption{Dimensions of the helices in each block of the lens. The radius of the wire $r_0=0.33$~mm.}
\label{tabl:table3}
\end{table}
%The absorber used in the 3 layers structure consists of 29 unit cells of equal dimensions, the inclusions are made out of ``0.1mm radius'' copper wire. The pitch is 1.11mm, the helix radius 1.71mm, and pitch angle is $5.9\degree$.\\
%The metamirror consists of 6 $\Omega$ inclusions repeated 5 times to fit the proposed 3 layer structure to preform the reflecting task. The last instance lacks one inclusion as we only have 29 unit cells for the transmit array and the absorber. Two types of $\Omega$ inclusions are used and shown in the figure below.\\

\section*{Acknowledgment}

This work was supported by Academy of Finland (project 287894). The authors would like to thank Dr. Y. Ra'di for his help and contribution during the experimental phase.

\ifCLASSOPTIONcaptionsoff
  \newpage
\fi

\end{document}